\def\q{{\bm q}}
\def\x{{\bm x}}

\def\tr{\operatorname{tr}}

\def\Im{\operatorname{Im}}

\def\gSG{g_{\rm SG}}
\def\pol{\bar\varepsilon}
\def\kbig{\bar k}

\def\Aamp{{\cal N}_A}
\def\Acl{A_{\rm cl}}

\def\formL{\Psi}

\def\calGRup{{\cal G}^{\rm R}}

\def\uB{u_{\rm B}}
\def\zB{z_{\rm B}}

\def\Field{{\cal A}}

\def\Prob{\operatorname{Prob}}
\def\del{\epsilon}
\def\xz{x^{\mathsf{3}}}
\def\xo{x^{\mathsf{0}}}
\def\five{{\mathsf{5}}}
\def\xstop{\xz_{\rm stop}}
\def\xtypical{\xz_{\rm typical}}
\def\xmax{\xz_{\rm max}}
\def\xo{x^{\mathsf{0}}}

\documentclass[prd,preprint,eqsecnum,nofootinbib,amsmath,amssymb,
               tightenlines,dvips]{revtex4}
\usepackage{graphicx}
\usepackage{bm}
\usepackage{amsfonts}

\begin {document}



\title
    {
      Jet quenching in hot strongly coupled gauge theories
      simplified
    }

\author{
  Peter Arnold and Diana Vaman
}
\affiliation
    {%
    Department of Physics,
    University of Virginia, Box 400714,
    Charlottesville, Virginia 22904, USA
    }%

\date {\today}

\begin {abstract}%
{%
  Theoretical studies of jet stopping in strongly-coupled QCD-like
  plasmas have used gauge-gravity duality to find
  that the maximum stopping distance scales
  like $E^{1/3}$ for large jet energies $E$.
  In recent work studying jets that are created by
  finite-size sources in the gauge theory, we found an additional scale:
  the typical (as opposed to maximum) jet
  stopping distance scales like $(EL)^{1/4}$, where
  $L$ is the size of the space-time region where the jet is
  created.  In this paper, we show that the results of our
  previous, somewhat involved computation in the gravity dual,
  and the $(EL)^{1/4}$ scale in particular,
  can be very easily
  reproduced and understood
  in terms of the distance that
  high-energy particles travel in
  AdS$_5$-Schwarzschild space before falling into the black brane.
  We also investigate how stopping distances depend on the conformal
  dimension of the source operator used to create the jet.
}%
\end {abstract}

\maketitle
\thispagestyle {empty}


\section {Introduction and Results}
\label{sec:intro}

\begin {figure}
\begin {center}
  \includegraphics[scale=0.4]{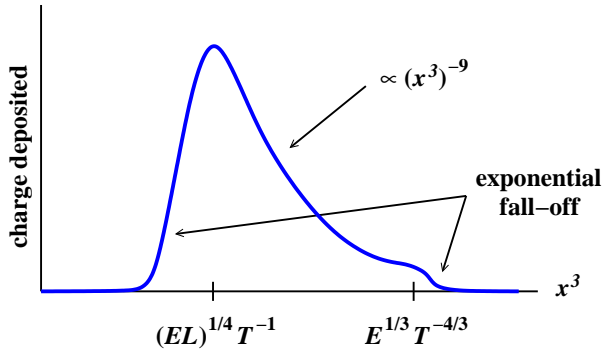}
  \caption{
     \label{fig:stop}
     The average deposition of charge as a function of $\xz$ for
     jets created by the source described in Sec.\ \ref{sec:PR}
     and in Ref.\ \cite{adsjet}.
  }
\end {center}
\end {figure}

Various authors \cite{GubserGluon,HIM,CheslerQuark,adsjet}
have made use of gauge-gravity duality to
study the stopping distance of massless, high-energy
jets in a strongly-coupled plasma of ${\cal N}{=}4$ supersymmetric
Yang Mills theory (with and without the addition of
fundamental-charge matter).  All have found that the furthest that
such a jet penetrates through the plasma scales with energy as
$E^{1/3}$.  Most of these methods specified the initial conditions
of the problem in the gravity description of the problem, and it
is not completely clear exactly what these initial conditions correspond
to in the gauge theory.
However, we recently showed \cite{adsjet}
one possible way to set up the entire problem
directly in the gauge theory, only then translating to the gravity
description using the conventional elements of the AdS/CFT dictionary.
We specifically studied jets that carried R charge, and we measured
how far that charge traveled before stopping and thermalizing.
Though we did find that the furthest charge would travel through
the plasma scaled as $E^{1/3}$, we also found that, on average, almost all
of our jet's charge stopped and thermalized at a shorter distance
that scales as $(EL)^{1/4}$, where $L$ is the size of the space-time
region where our jet was created.  Fig.\ \ref{fig:stop} shows
a qualitative picture of our result for, on average, how much of our
jet's charge was deposited as a function of distance traveled $\xz$.
(Our convention here is to write 4-dimensional space-time position as
$x^\mu$ and take our jets to be created near the origin, traveling
in the $\xz$ direction.)
Between the $(EL)^{1/4}$ scale and the $E^{1/3}$ scale, the
distribution falls algebraically like $(\xz)^{-9}$ for jets created
by the source used in Ref.\ \cite{adsjet}.
We will work in units where
$2\pi T = 1$, and in those units the specific
formula we derived for Fig.\ \ref{fig:stop} was
\begin {subequations}
\label {eq:final}
\begin {align}
   \Prob(\xz) &\simeq
   2
      \frac{(4 c^4 EL)^2}{(2\xz)^9} \,
      \formL\Bigl(-\frac{c^4 EL}{(2\xz)^4}\Bigr)
   &\mbox{for $\xz \ll E^{1/3}$\phantom{,}}
\label {eq:final1}
\\
\intertext{and}
   \Prob(\xz) &\simeq
      4 \frac{(c_2 L)^2}{E} \, \formL(0) \,
      \exp\left( -\frac{2 c_1 \xz}{E^{1/3}} \right)
   &\mbox{for $\xz \gg E^{1/3}$,}
\label {eq:final2}
\end {align}
\end {subequations}
where
$\formL(y)$ is a source-dependent function that suppresses
$|y| \gg 1$, causing suppression of $\xz \ll (E L)^{1/4}$
above.  The $c$'s are constants given by
\begin {equation}
   c \equiv \frac{\Gamma^2(\tfrac14)}{(2\pi)^{1/2}} \,,
\label {eq:num}
\end {equation}
$c_1 \simeq \, 0.927 \,$, and $c_2 \simeq 3.2 \,$.

The calculation that produced (\ref{eq:final}) was long and not
particularly enlightening as to the origin of the $(EL)^{1/4}$
scale.  The purpose of the current paper is to show how that
scale, and then the precise result (\ref{eq:final1})
for the case $\xz \ll E^{1/3}$, can be derived from a very simple
calculation of how far a classical massless particle
travels in AdS$_5$-Schwarzschild space before falling into the
black brane.  In the process, we will learn more about exactly what
feature of the source determines the $(EL)^{1/4}$ scale.
We will see that it is
not directly the size 
but the typical ``virtuality'' $q^2$
of the source that matters (where $q^2 \equiv q_\mu \eta^{\mu\nu} q_\nu$
is squared 4-momentum).

Our analysis of the distance traveled by
falling particles in AdS$_5$-Schwarzschild will be
essentially the same as an earlier analysis by
Gubser et al.\ \cite{GubserGluon} and
Chesler et al.\ \cite{CheslerQuark}, who used it in a discussion
of the falling endpoint of a classical string.
The difference here will be one of context and application:
Our analysis of jets \cite{adsjet} does not involve
classical strings, and we will use the falling particles to
explain the $(EL)^{1/4}$ scale.

\begin {figure}
\begin {center}
  \includegraphics[scale=0.3]{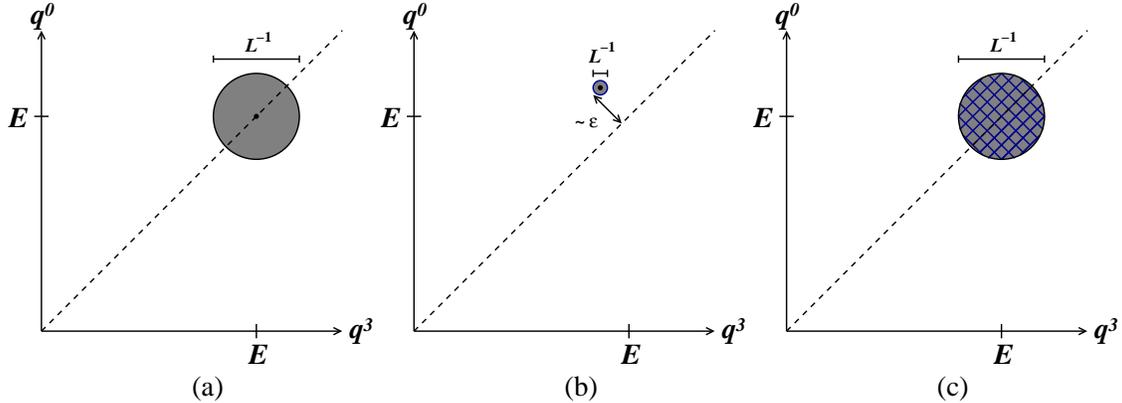}
  \caption{
     \label{fig:source}
     Qualitative picture of
     momenta contributing to the source (\ref{eq:source1})
     used to generate jets
     (a) for the calculation originally used to find (\ref{eq:final}),
     with $L^{-1} \ll E$,
     and (b) in the case $L^{-1} \ll \del \ll E$ of (\ref{eq:kdelta}).
     Figure (c) depicts (a) as a superposition of distributions
     of type (b).  The cells in (c)
     that are extremely close to the
     light cone cannot be treated in particle approximation, but
     the contribution of cells that can be treated so dominates when
     $L \ll$ the maximum stopping distance
     scale $E^{1/3}$.
  }
\end {center}
\end {figure}

In our earlier work \cite{adsjet}, we created the jet by
turning on a small-amplitude source whose space-time dependence had
the form
\begin {equation}
  \operatorname{source}(x) \sim
  e^{i \bar k\cdot x} \, \Lambda_L(x)
\label {eq:source1}
\end {equation}
of (i) a high-energy plane wave $e^{i \bar k\cdot x}$ times 
(ii) a slowly
varying envelope function $\Lambda_L(x)$ that localizes the
source to within a distance $L$ of the origin in both space and
time.  We took $\bar k$ to be light-like:
\begin {equation}
   {\bar k}^\mu = (E,0,0,E) .
\label {eq:kbig}
\end {equation}
In addition, for the sake of simplicity, we took the source to be
translation invariant in the two transverse directions.  So, for example,
\begin {equation}
   \Lambda_L(x)
   = e^{-\frac12 (x^0/L)^2} e^{-\frac12 (\xz/L)^2} .
\label {eq:Genvelope}
\end {equation}
The Fourier transform of the source (\ref{eq:source1}) is non-negligible
in the region of momentum space
depicted in Fig.\ \ref{fig:source}a: a region centered
on $\bar k$ with width of order $L^{-1}$.
We take $L^{-1} \ll E$.  Note that this source covers
a range of values of $q^2$, from 0 to order $\pm E/L$, and
the typical size of $|q^2|$ is order $E/L$.

In the gravity description, this source causes a localized perturbation
on the boundary of AdS$_5$-Schwarzschild space-time, which then
propagates as a wave into the fifth dimension, eventually falling into
the black brane horizon.  The analysis of jet stopping in
Refs.\ \cite{HIM,adsjet} was based on the analysis of such
5-dimensional waves.

\begin {figure}
\begin {center}
  \includegraphics[scale=0.4]{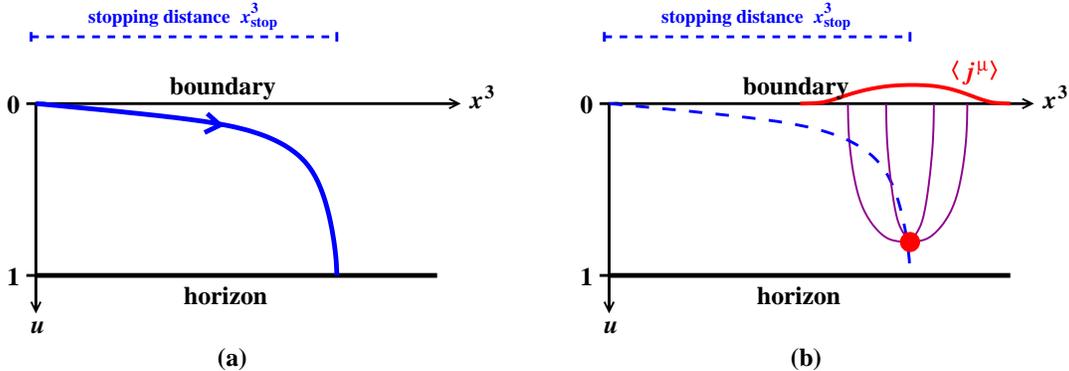}
  \caption{
     \label{fig:fall}
     (a) A classical particle in the AdS$_5$-Schwarzschild space-time,
     moving in the $\xz$ direction as it falls from the boundary
     to the black brane in the fifth dimension $u$.
     (b) The presence of the particle (the large dot)
     perturbs the boundary theory
     in a manner that spreads out diffusively as the particle
     approaches the horizon for $\xo \to \infty$.
  }
\end {center}
\end {figure}

Now imagine instead a source where $\bar k$ is slightly off the
light-cone,
\begin {equation}
   {\bar k}^\mu = (E+\del,0,0,E-\del)
\label {eq:kdelta}
\end {equation}
with $\del \ll E$,
and where the envelope size $L$ is wide enough that the picture in momentum
space looks like Fig.\ \ref{fig:source}b instead of Fig.\
\ref{fig:source}a,
with the spread $1/L$ in momenta small compared to $\del$.
In this case, the $q^2$ of the source is approximately well defined,
with $q^2 \simeq \bar k^2 \simeq -4 E \del$.
We will show that in this case the wave created by the boundary
perturbation is localized into a small wave packet, whose motion
may be approximated by that of a classical, massless particle
which starts at the boundary, traveling in the $\xz$ direction,
with 4-momentum proportional to $q$.
The trajectory of such a particle is shown qualitatively in
Fig.\ \ref{fig:fall}a.
By a simple calculation, we will find that the
particle falls into the horizon after covering a distance
\begin {equation}
   \xstop
   \simeq \frac{c}{\sqrt2} \left( \frac{|\q|^2}{-q^2} \right)^{1/4}
   \simeq \frac{c}{2} \left( \frac{E}{\del} \right)^{1/4}
   .
\label {eq:stop}
\end {equation}
where the constant $c$ is given by (\ref{eq:num}).
As measured by boundary time $\xo$, the particle takes an
infinite amount of time to fall into the horizon.  As it gets
closer and closer to the horizon, the boundary distortion
that the particle creates (see Fig.\ \ref{fig:fall}b) becomes weaker and
more spread out, which corresponds to charge diffusion in the
boundary theory after the jet stops and thermalizes.
This qualitative picture is similar to the qualitative
picture of the effects of a classical string falling into
the horizon given in Refs.\ \cite{GubserGluon,CJK}.%
\footnote{
  See in particular the discussion surrounding Fig.\ 2 of
  Ref.\ \cite{CJK}, which inspired our Fig.\ \ref{fig:fall}b.
  See also Ref.\ \cite{GubserPointlike}.
}

Note that the stopping distance (\ref{eq:stop}) only makes sense for
$q^2 < 0$ (i.e.\ $\del > 0$).
The $q^2 {>} 0$  components of a source do not create an excitation of the
system that persists after the source turns off
and so are not relevant \cite{HIM,adsjet}.

Now consider the original source of Fig.\ \ref{fig:source}a as
a superposition of sources like Fig.\ \ref{fig:source}b, as depicted
in Fig.\ \ref{fig:source}c.  Since sources with different values of
$\del$ have different stopping distances (\ref{eq:stop}), we might
guess that the different pieces of this superposition do not
interfere and so the source of Fig.\ \ref{fig:source}a simply
produces a distribution of stopping distances, weighted by independent
probabilities that the source produces a jet with a particular
$q^2$.  That is,
\begin {equation}
   \Prob(\xz) \simeq
   \int d(q^2) \, {\cal P}(q^2) \,
   \delta\bigl(\xz - \xstop(q^2)\bigr) ,
\label {eq:Pstop}
\end {equation}
where ${\cal P}(q^2)$ is the probability density for the source
to produce a
jet with a given $q^2$, and where the stopping distance
$\xstop(q^2)$ is given by
(\ref{eq:stop}).  We will verify that this formula
precisely reproduces the $\xz \ll E^{1/3}$ case (\ref{eq:final1})
of our previous result.

We can now see where the $(EL)^{1/4}$ scale comes from.
It is only time-like source momenta $q^2 < 0$ that produce
jets.  The typical value of time-like $q^2$ for the source of
Fig.\ \ref{fig:source}a is $q^2 \sim -E/L$, corresponding to
$\del \sim L^{-1}$.  Putting this into
(\ref{eq:stop}), the typical stopping distance in this
case is therefore
\begin {equation}
   \xtypical \sim (EL)^{1/4} \,.
\label {eq:ELfourth}
\end {equation}
Note that it is the $q^2$ of the source that determines the
stopping distance, and that the typical value of $q^2$ is determined
by $L$ in the case of
Fig.\ \ref{fig:source}a.

The estimate (\ref{eq:ELfourth}) of the stopping distance ceases
to make sense if the size $L$ of the source becomes as large
as the stopping distance itself.  This happens when
\begin {equation}
   L \sim \xstop \sim (EL)^{1/4} ,
\end {equation}
which gives
\begin {equation}
   \xstop \sim E^{1/3} .
\end {equation}
We will see later that this is precisely the case where the
wave packet in AdS$_5$-Schwarzschild
can no longer be approximated as a particle.
The moral is that the simple particle picture gives us not only
the $(EL)^{1/4}$
scale but also, simply by estimating where it breaks down, the
$E^{1/3}$ scale as well.

\bigskip

\begin {figure}
\begin {center}
  \includegraphics[scale=0.4]{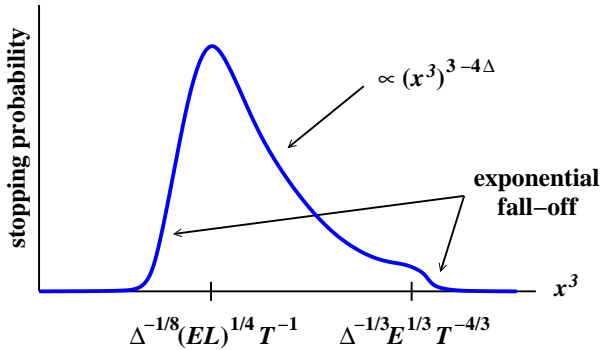}
  \caption{
     \label{fig:stopDelta}
     The probability distribution of jet stopping distances for
     scalar or transverse BPS sources with conformal dimension
     $\Delta$.
     Fig.\ \ref{fig:stop}
     corresponds to $\Delta{=}3$.
     The scales
     $\xtypical \sim (EL/\sqrt{\Delta})^{1/4}$ and
     $\xmax \sim (E/\Delta)^{1/3}$ indicated
     along the vertical axis assume that $\Delta$ is held fixed when
     taking the limit of large energy $E$ (as well as
     large coupling $g^2 N_{\rm c}$ and large $N_{\rm c}$).
     The parametric scaling with $\Delta$ indicated for $\xtypical$
     assumes a Gaussian source envelope (\ref{eq:Genvelope}), but the
     other features shown in the figure are independent of the details
     of the source envelope.
  }
\end {center}
\end {figure}

In the next section, we will briefly review the trajectories of
massless particles in AdS$_5$-Schwarzschild and derive the
corresponding stopping distance (\ref{eq:stop}).
In section \ref{sec:waves}, we discuss the conditions for
being able to approximate the 5-dimensional wave problem with
particle trajectories and verify that they apply in the
case of interest.  Then we use the particle picture in
section \ref{sec:reproduce} to simply reproduce our original
result (\ref{eq:final1}) for charge deposition for
$\xz \ll E^{1/3}$.
In section \ref{sec:massive}, we generalize our results to jets
created by other types of source operators than those originally
considered in Ref.\ \cite{adsjet}.  We will see that
Fig.\ \ref{fig:stop} is modified to Fig.\ \ref{fig:stopDelta}.
Finally, we offer our conclusions
in section \ref{sec:conclusion}.

As an aside, some readers may be curious how the stopping distance
scales $(EL)^{1/4}$ and $E^{1/3}$ generalize to other dimensions.
On the gravity side, it is very easy to generalize the results
of this paper to
different space-time dimensions $d$ of the boundary,
but it is less certain what strongly coupled
field theories these classical gravity theories correspond to.
(See Refs.\ \cite{DifferentDims} for proposals.)
Ignoring the question of interpretation, we show in
Appendix \ref{app:dim} that $(EL)^{1/4}$ and
$E^{1/3}$ generalize to
$(EL)^{(d-2)/2d}$ and $E^{(d-2)/(d+2)}$ respectively
for $d>2$.

In this paper, we will use the convention that Greek indices
run over the 4 space-time
dimensions ($\mu= 0,1,2,3$) of the boundary theory and
capital roman indices run over all five dimensions
($I=0,1,2,3,5$) of the AdS$_5$-Schwarzschild space-time.
The symbol $q^2$ will refer to the squared 4-momentum,
$q^2 \equiv q_\mu \eta^{\mu\nu} q_\nu = -\omega^2+|\q|^2$.
When we use light-cone coordinates, our conventions will be
\begin {equation}
   V^\pm \equiv V^{\textsf{3}} \pm V^{\textsf{0}} ,
   \qquad
   V_\pm \equiv \tfrac12 V^\mp
   = \tfrac12 (V^{\textsf{3}} \mp V^{\textsf{0}})
\end {equation}
for any 4-vector $V$.
Throughout this paper. the adjective ``transverse'' will refer to
the {\it spatial}\/ directions 1 and 2 orthogonal to $\q$.%
\footnote{
  This is as opposed to the alternative usage of ``transverse''
  to mean all three
  space-time components orthogonal to $q_\mu$.
  Note that the spatial directions transverse to $\q$ are the
  same for all $q$ in
  our problem because we take our source (\ref{eq:source})
  to be translationally
  invariant in the transverse directions.
}


\section {Review of Falling Massless Particles}

Null geodesics in a 5-dimensional space with 4-dimensional translation
invariance are given by (see Appendix \ref{app:null}):
\begin {equation}
   x^\mu(x^\five) = \int \sqrt{g_{\five\five}} \, dx^\five \> 
      \frac{ g^{\mu\nu} q_\nu }
           { (-q_\alpha g^{\alpha\beta} q_\beta)^{1/2} }
   \,,
\label {eq:null}
\end {equation}
where $g$ is the 5-dimensional metric and $q_I$ is a constant of
motion for $I=0,1,2,3$.
We will work in coordinates where the metric is%
\footnote{
   Our formulas in this paper would be a little tidier (fewer square roots)
   if we worked with the coordinate $z \equiv 2 \sqrt{u}$
   instead of $u$.  We will stick with $u$ in order to facilitate
   comparison with our previous work \cite{adsjet}.
}
\begin {equation}
  ds^2 = \frac{R^2}{4} \left[ \frac{1}{u}(-f \, dt^2 + d\x^2)
         + \frac{1}{u^2 f} \, du^2 \right] ,
\label {eq:metric}
\end {equation}
where $f\equiv 1-u^2$, and $R$ is the AdS$_5$ radius.
The boundary is at $u{=}0$ and the horizon at $u{=}1$.
If we take the 3-momentum $\q$ to point in the $\xz$ direction,
writing $q_\mu = (-\omega,0,0,|\q|)$, then (\ref{eq:null}) gives
the total distance $\xz$ traveled in falling from the boundary
to the horizon to be
\begin {equation}
   \xstop = \int_0^1 \frac{du}{\sqrt{u(u^2-\frac{q^2}{|\q|^2})}} \,,
\label {eq:stop1}
\end {equation}
where $q^2 \equiv q_\mu \eta^{\mu\nu} q_\nu$ is the flat-space
square of the 4-momentum.
This is the same result as Refs.\ \cite{GubserGluon,CheslerQuark}.%
\footnote{
  Our (\ref{eq:intdx3}) corresponds to the first part of Eq.\ (5.3)
  of Gubser et al.\ \cite{GubserGluon}, where their $y$ is our
  $\sqrt{u}$, their $z_H$ is $2$ in our units $2\pi T{=}1$, their $p_1/p_0$
  is replaced by the $q_3/q_0$ of the momentum $q_\mu$ typical of our
  source, and their $y_{\rm UV}$ is set to zero.
  It also corresponds to Eq.\ (4.28) of Chesler et al.\
  \cite{CheslerQuark}, where their $u$ is our $2\sqrt{u}$, their
  $u_h=2$ in our units, their $\xi$ is replaced by our
  $q_0/q_1$, and their $u_*$ is set to zero.
}

Now let us apply this result to the case $|q^2| \ll |\q|^2 \simeq E^2$
relevant to the source of Fig.\ \ref{fig:source}b.  
For small $-q^2$, the integral of (\ref{eq:stop1}) is dominated by
small $u$, and so we may approximate
\begin {equation}
   \xstop
   \simeq \int_0^\infty \frac{du}{\sqrt{u(u^2-\frac{q^2}{|\q|^2})}}
   = \frac{c}{\sqrt2} \left( \frac{|\q|^2}{-q^2} \right)^{1/4} ,
\label {eq:intdx3}
\end {equation}
which gives (\ref{eq:stop}).


\section {Wave Packets and Geometric Optics}
\label {sec:waves}

\begin {figure}
\begin {center}
  \includegraphics[scale=0.5]{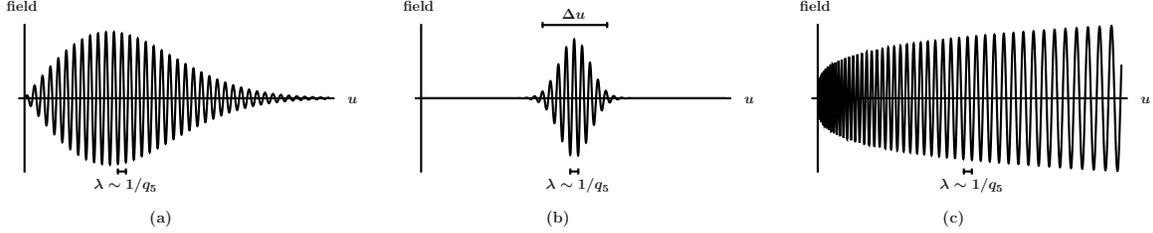}
  \caption{
     \label{fig:packet}
  (a,b) A snapshot in time $\xo$ of waves in the fifth dimension $u$ for
  times after the boundary source has turned off but early enough that
  $u \lesssim u_\star \ll 1$ (that is, before the wave gets very close to
  the horizon).
  (a) shows the type of wave generated by a localized
  source that superposes a range of $q^2$ values such as Fig.\
  \ref{fig:source}a.
  (b) shows the wave packet generated by a source with approximately
  well-defined $q^2$ such as Fig.\ \ref{fig:source}b.
  (c) shows a single 4-momentum component, corresponding to a single, definite
  value of 4-momentum $q_\mu$.
  Case (a) matches Fig.\ 7a of our earlier paper \cite{adsjet}.
  } 
\end {center}
\end {figure}

In this section, we will discuss the conditions necessary for making
the particle approximation.  A wave packet behaves like a particle
when it is wide enough to contain many phase oscillations of the field
yet small enough that the properties of the background do not vary
significantly across its width, as depicted in Fig.\ \ref{fig:packet}b
at a particular moment in time.  We can arrange such a width
provided the background properties do not vary significantly over
one wavelength of the phase oscillation.  This is the
geometric optics limit, which we referred to in our earlier
work \cite{adsjet} as a WKB approximation.
To check the geometric optics limit, one may focus as in
Fig.\ \ref{fig:packet}c on a wave
with a single, generic value of $q$ typical of the wave packet,
and investigate how much things change over one phase oscillation.

To assess whether a wave packet is adequately particle-like to use
a particle-based calculation of the stopping distance, it will be
helpful to understand what the important scale for $u$ is in
determining the stopping distance (\ref{eq:intdx3}).
The integral in (\ref{eq:intdx3})
is dominated by $u$ of order
\begin {equation}
   u_\star \sim \sqrt{\frac{-q^2}{|\q|^2}}
   \sim \sqrt{\frac{\del}{E}} .
\label {eq:ustar}
\end {equation}
The relevant question is then whether the background varies
significantly across one phase oscillation for $u \sim u_\star$.
By (\ref{eq:stop}),
the distances $\xstop \lesssim E^{1/3}$ relevant to jet stopping
correspond to $\del \gtrsim E^{-1/3}$ in the particle picture and
so to $u_\star \lesssim E^{-2/3} \ll 1$.  So we may focus on small
$u$ in what follows.

\subsection {Geometric Optics}

For a massless 5-dimensional field with definite 4-momentum $q_\mu$,
the exponential in the
WKB approximation is
\begin {equation}
   \exp\left(i q_\mu x^\mu + i \int dx^\five \, q_\five(x^\five) \right) ,
\label {eq:wave}
\end {equation}
where the $q_\mu$ are constant and $q_\five(x^\five)$ is determined by
the 5-dimensional massless condition $q_I g^{IJ} q_J = 0$,
giving%
\footnote{
  The null geodesics (\ref{eq:null}) can be expressed
  in terms of $q_\five$ as
  $x^\mu(x^\five) = - \int dx^\five (\partial q_\five/\partial q_\mu)$.
  This particle formula is simply the saddle point condition
  $\partial_{q_\mu} [ i q_\nu x^\nu + i \int dx^\five\> q_\five(x^\five)] = 0$
  with respect to $q_\mu$ for the wave (\ref{eq:wave}).
  Also, $\int dx^\five \>q_\five(x^\five)$ was referred to as
  the WKB exponent $S$ in Ref.\ \cite{adsjet}, where
  various expansions of the integral may be found.
}
\begin {equation}
   q_\five(x^\five) =
   \sqrt{g_{\five\five}(-q_\mu \, g^{\mu\nu} \, q_\nu)}
   .
\end {equation}
For the metric (\ref{eq:metric}), this is
\begin {equation}
   q_\five(u) =
   \frac{1}{f} \sqrt{ \frac{u^2|\q|^2-q^2}{u} } .
\label {eq:q5u}
\end {equation}
For the geometric optics limit, we need the wavelength to vary
insignificantly over one wavelength.%
\footnote{
   This condition can be phrased in an
   $x^\five$-reparametrization invariant way as
   $\nabla_\five (1/q_\five) \ll 1$, which is
   $(g_{\five\five})^{-1/2} \partial_\five
    [(g_{\five\five})^{1/2} / q_\five] \ll 1$.
   We give more detail in Appendix \ref{app:GO},
   in the specific context of the particular type of source operator that
   we used in our original calculation.
}
For the important values $u \sim u_\star \ll 1$ of $u$, the
wavelength $\lambda(u) \sim 1/q_\five(u)$ in the fifth dimension
satisfies this condition if
\begin {equation}
   u_\star \, q_\five(u_\star) \gg 1 ,
\end {equation}
which, using (\ref{eq:q5u}), is%
\footnote{
  In the language of Ref.\ \cite{adsjet}, the condition
  $u_\star \gg 1/(-q^2)$ is
  $u_\star \gg u_{\rm match}$.
}
\begin {equation}
  u_\star \gg \frac{1}{-q^2} .
\label {eq:GOconditionu}
\end {equation}
Using the size (\ref{eq:ustar}) of $u_\star$, this condition is
$(-q^2)^3/|\q|^2 \gg 1$, or equivalently
\begin {equation}
   \del \gg E^{-1/3} .
\end {equation}
Referring to the stopping distance (\ref{eq:stop}), we then see that the
approximation of geometric optics, necessary for a particle interpretation,
breaks down unless
\begin {equation}
   \xstop \ll E^{1/3} .
\label {eq:GOcondition}
\end {equation}
Within the context of our approach to jet stopping,
a proper analysis of what happens at distances $\gtrsim E^{1/3}$
requires a wave rather than particle description of the problem, as in
Refs.\ \cite{HIM,adsjet}.
The wave analysis gives exponential fall-off for propagation beyond
$E^{1/3}$, as described by (\ref{eq:final2}).

More on the geometric optics approximation can be found in
Appendix \ref{app:GO}.

\subsection {Wave Packets}

The geometric optics limit for (\ref{eq:GOcondition}) allows us to
create localized wave packets.  Here we will see how wide those
wave packets are in $u$ for sources of the form of Fig.\ \ref{fig:source}b.
We are primarily interested in the case where the center of the wave packet
is at the critical scale $u \sim u_\star$ in the fifth dimension.
However,
the presentation will be a little more straightforward if we
first  make parametric
estimates for earlier times, when the center of the wave packet is
at $u \ll u_\star$, and then extrapolate those parametric estimates to
$u \sim u_\star$.

For $u \ll u_\star$, the distance traveled (\ref{eq:null}) for
a massless particle is
\begin {equation}
   \xz(u) = \int_0^u \frac{du'}{\sqrt{u'(u'^2-\frac{q^2}{|\q|^2})}}
   \simeq 2\sqrt{ \frac{u |\q|^2}{-q^2} }
   \simeq \sqrt{\frac{u E}{\del}} \,.
\end {equation}
Turning this around, the location of the particle in the
fifth dimension is
\begin {equation}
   u \simeq \frac{\del \, (\xz)^2}{E} \,.
\label {eq:usmall}
\end {equation}
When the particle is replaced by a wave packet, there are two sources
of uncertainty.  The size $L$ of the source introduces an uncertainty
in the initial position of the excitation of $\Delta\xz \sim L$.
It also introduces an uncertainty in the 4-momentum $q$, and
so in $\del$, of $\Delta\del \sim 1/L$ as in Fig.\ \ref{fig:source}b.
From (\ref{eq:usmall}), the combined uncertainty $\Delta u$
in $u$ is then of order
\begin {equation}
   \frac{\Delta u}{u} \sim
   \max\left(
     \frac{\Delta\del}{\del}
     \> , \>
     \frac{\Delta\xz}{\xz}
   \right)
   \sim
   \max\left(
     \frac{1}{L \del}
     \> , \>
     L\sqrt{\frac{\del}{uE}}
   \right) .
\label {eq:Du}
\end {equation}

Extrapolating this parametric estimate to the case $u\sim u_\star$ of
interest, (\ref{eq:ustar}) and (\ref{eq:Du}) give
\begin {equation}
   \left(\frac{\Delta u}{u}\right)_\star
   \sim \max\left(
     \frac{1}{L\del}
     \> , \>
     L\left(\frac{\del}{E}\right)^{1/4}
   \right) .
\end {equation}
The wave packet will then be localized provided
(i) $L \gg 1/\del$, as in Fig.\ \ref{fig:source}b,
and (ii) $L \ll (E/\del)^{1/4}$.  By (\ref{eq:stop}),
the last condition
is just the condition that $L$ be much less than the
stopping distance $\xstop$.


\section {Reproducing the distribution of stopping distances}
\label {sec:reproduce}

Now we will show that the formula (\ref{eq:final1}) found in
our earlier work \cite{adsjet} for
the average distribution of charge deposition
can be understood
as a convolution (\ref{eq:Pstop}) of the particle stopping distance
with the probability density ${\cal P}(q^2)$ for the source
to create a jet with a given $q^2$.

We will later give in section \ref{sec:massive} a very general
argument, based on dimensional analysis, for determining
the ${\cal P}(q^2)$ associated with different choices of source
operator.  This argument will also require a discussion of massive
fields and massive particles in the gravity dual.
For the moment, we will be less general, stick to the specific
type of source operator that we used in previous work, and
show how to extract ${\cal P}(q^2)$ from a result for the
average total charge produced by the operator.  Readers who
would prefer to just see the more general argument may skip
section \ref{sec:PR} below and instead wait for section
\ref{sec:Pm}.


\subsection {Extracting \boldmath${\cal P}(q^2)$ from results in
             the literature}
\label {sec:PR}

In Ref.\ \cite{adsjet}, we used a source involving R-current
operators $j^a_\mu$.
Specifically, we modified the 4-dimensional gauge theory Lagrangian by
\begin {equation}
   {\cal L} \to {\cal L} + j_\mu^a \Acl^{a\mu} ,
\label {eq:Lsource}
\end {equation}
with a localized background field
\begin {equation}
   \Acl^{\mu}(x)
   = \pol^\mu \Aamp \Bigl[
       \frac{\tau^+}{2} \, e^{i \kbig\cdot x} +
       {\rm h.c.}
      \Bigr] \, \Lambda_L(x) ,
\label{eq:source}
\end {equation}
where
$\Aamp$ is an arbitrarily small source amplitude,
$\pol$ is a transverse linear polarization,
and $\tau^i$ are Pauli matrices
for any SU(2) subgroup of the
SU(4) R-symmetry.
We then measured the response $\langle j^{(3)\mu}(x)\rangle$ of the
R charge current associated with $\tau^3/2$.
The gravity dual to the R charge current operators is a
massless 5-dimensional SU(4) gauge field.
We chose our source (\ref{eq:source}) to be translationally invariant
in the transverse spatial directions ($x^1,x^2$) to simplify the
calculation.
In what follows, we will refer to the R charge associated with
$\tau^3/2$ as simply ``the charge.''

For an arbitrarily small source amplitude $\Aamp$,
the source will usually have no effect at all on the system.
On rare occasions, with probability proportional to $\Aamp^2$,
the source creates an excitation (in our case a ``jet'') with the same quantum
numbers as the source operator.  In our case (\ref{eq:source}), that means it
creates a jet with total charge equal to 1.
The creation of an excitation with different quantum numbers would
be even higher-order in $\Aamp$ and so negligible.
Since an excitation (if any is created) has charge 1, the
{\it average} total charge ${\cal Q}$ created by the source
is then equal to the probability of the source creating a jet.
We will see that from the previously calculated
result for ${\cal Q}$ we can then extract the probability
density ${\cal P}(q^2)$.

In Ref.\ \cite{adsjet},%
\footnote{
  See specifically Appendix A of Ref.\ \cite{adsjet}.
}
we showed how to use the field theory
Ward identity to make a simple calculation of the
average charge ${\cal Q}$.
Here we will just quote the result, which was
\begin {equation}
   {\cal Q}
   \simeq
   \frac{2\pi \Aamp^2}{\gSG^2} \int \frac{d^4q}{(2\pi)^4}
   \> \theta(-q^2) \, |q^2|
   \bigl|\tilde\Lambda_L(q-\bar k)\bigr|^2 ,
\label {eq:charge1}
\end {equation}
where $\gSG^2 = 4\pi/N_{\rm c}$ ($N_{\rm c}{\to}\infty$ is the
number of colors)
and $\theta$ is the step function.
The $\tilde\Lambda_L(q-\bar k)$ in
(\ref{eq:charge1}) is simply the Fourier transform of the
$x$-dependence (\ref{eq:source1}) of the source.
Our source only has support for $q = \bar k + \Delta q$ with
$\Delta q$ small compared to $\bar k$,
in which case $q^2 \simeq 4Eq_+$.
Eq.\ (\ref{eq:charge1}) may then be approximated as
\begin {equation}
   {\cal Q}
   \simeq
   \frac{8\pi E\Aamp^2}{\gSG^2} V_\perp
   \int \frac{2 dq_+ \, d\Delta q_-}{(2\pi)^2}
   \> \theta(-q_+) \, |q_+|
   \bigl|\tilde\Lambda_L^{(2)}(q_+,\Delta q_-)\bigr|^2 ,
\label {eq:charge}
\end {equation}
where $\tilde\Lambda_L^{(2)}$ is the two-dimensional Fourier transform
\begin {equation}
   \tilde\Lambda_L^{(2)}(q_+,q_-) =
   \int \frac{dx^+\,dx^-}{2} \> \Lambda_L(x) \,
   e^{-i(q_+ x^+ + q_- x^-)}
\label {eq:envelope2}
\end {equation}
of the source envelope and $V_\perp$ is the area of transverse
space ($x^1$ and $x^2$).
Since (\ref{eq:charge}) involves an integral over
$q_+$, and since ${\cal Q}$ is the total probability of creating a jet,
it is natural to interpret (\ref{eq:charge}) as
giving a probability density
\begin {equation}
   P_+(q_+)
   \simeq
   \frac{16\pi E\Aamp^2}{\gSG^2} V_\perp
   \int \frac{d\Delta q_-}{(2\pi)^2}
   \> \theta(-q_+) \, |q_+|
   \bigl|\tilde\Lambda_L^{(2)}(q_+,\Delta q_-)\bigr|^2
\end {equation}
for producing a jet with a given value of $q_+$.
If a jet
{\it is}\/ produced, the probability distribution for its
$q_+$ is then the relative probability
\begin {equation}
   {\cal P}_+(\bar q_+) \equiv
   \frac{P_+(\bar q_+)}{\cal Q}
   \simeq
   \frac{
     \theta(-\bar q_+) \, |\bar q_+|
     \int d\Delta q_- \>
     \bigl|\tilde\Lambda_L^{(2)}(\bar q_+,\Delta q_-)\bigr|^2
   }{
     \int dq_+ \> d\Delta q_- \>
     \theta(-q_+) \, |q_+|
     \bigl|\tilde\Lambda_L^{(2)}(q_+,\Delta q_-)\bigr|^2
   } \,.
\label {eq:Pplus0}
\end {equation}
Here we've put a bar over the argument of $P_+$ just
to distinguish it from the $q_+$ integration variable
in the denominator on the right-hand side.
Now package the source dependence into the definition
\begin {equation}
   \Psi(\bar q_+ L) \equiv
   \frac{\int dq_- \> |\tilde\Lambda_L^{(2)}(\bar q_+,q_-)|^2}
        {4L^2 \int dq_+ \> dq_- \> \theta(-q_+) \,
           |q_+|  \bigl|\tilde\Lambda_L^{(2)}(q_+,q_-)\bigr|^2}
   \,.
\end {equation}
This is the definition we made in our earlier work \cite{adsjet}
for the $\Psi$ that appears in the charge deposition result
(\ref{eq:final}).  With this definition,
\begin {equation}
   {\cal P}_+(q_+) =
   4 L^2 \, \theta(-q_+) \, |q_+| \,
   \Psi(q_+ L) .
\label {eq:calP+}
\end {equation}


\subsection {Using \boldmath${\cal P}(q^2)$ to get \boldmath$\Prob(\xz)$}

Since $q^2 \simeq 4 E q_+$, the probability
distribution ${\cal P}(q^2)$ for $q^2$ is related
to the probability distribution ${\cal P}_+(q_+)$ of
(\ref{eq:calP+}) for $q_+$ by
\begin {equation}
   {\cal P}(q^2) \simeq
   \frac{1}{4E} \,
   {\cal P}_+\Bigl(\frac{q^2}{4E}\Bigr) .
\end {equation}
However, at this point it will be easier to just stick with
$q_+$ and ${\cal P}_+$.

Note that in each cell of Fig.\ \ref{fig:source}c, the typical value
of $q_+$ is just what we have previously called $-\del$ for that
cell.  The distribution (\ref{eq:Pstop}) of stopping distances 
based upon the picture of massless falling particles in 5 dimensions can
then be written as
\begin {align}
   \Prob(\xz) &\simeq
   \int d\del \, {\cal P}_+(-\del) \,
   \delta\bigl(\xz - \xstop(\del)\bigr) ,
\nonumber\\
   &\simeq
   4 L^2
   \int_0^\infty d\del \, \del \, \Psi(-L\del) \,
   \delta\Bigl(\xz - \tfrac{c}{2} \left(\tfrac{E}{\del}\right)^{1/4}\Bigr)
\nonumber\\
   &=
   2 \, \frac{(4 c^4 EL)^2}{(2\xz)^9} \,
   \Psi\Bigl(- \frac{c^4 EL}{(2\xz)^4}\Bigr) .
\label {eq:Probx3}
\end {align}
As promised, in the case $\xz \ll E^{1/3}$ where we have argued that the
particle picture should work,
this exactly reproduces our earlier result (\ref{eq:final1})
that came from a full, much more complicated calculation.


\section {Massive particles in 5 dimensions}
\label {sec:massive}

In the preceding sections, we have assumed that the 5-dimensional bulk
field dual to the source which creates the jet is massless, such as
the 5-dimensional gauge field dual to R current operators.
One may wonder what results change if we choose different types
of source operators that are instead dual to massive bulk fields.
In this section, we will see that the basic qualitative picture of
Fig.\ \ref{fig:stop} of the distribution of stopping distances
remains the same, except that the {\it exponent}\/ of the
$(\xz)^{-9}$ power-law tail changes, depending on the conformal
dimension of the source operator.

When we wish to make contact with a particular example, we will for
simplicity restrict attention to scalar BPS (e.g.\ chiral
primary \cite{chiral primary}) operators.%
\footnote{
  Examples of scalar BPS operators include the Lagrangian density and
  the symmetrized trace
  $\tr(\phi_{(i_1} \phi_{i_2} \cdots \phi_{i_n)})$, where $\phi_1$,
  $\phi_2$, and $\phi_3$ are the three complex scalar fields of
  ${\cal N}{=}4$ supersymmetric Yang Mills.
}
In that case, the mass $m$
of the 5-dimensional field is related to the scaling dimension
$\Delta$ of the operator by \cite{Witten}
\begin {equation}
  (R m)^2 = \Delta(\Delta-d) ,
\end {equation}
where
$R$ is the radius of AdS$_5$, and $d{=}4$ is the dimension of ordinary
space-time.  The possible values of $\Delta$ are bounded below by
\begin {equation}
   \Delta \ge \frac{d}{2} \,.
\end {equation}

In what follows, we will hold $m$ fixed when we consider the limit
of large jet energy $E$.  However, it will be interesting to consider
the case $\Delta \gg 1$ (i.e.\ $Rm \gg 1$)
in addition to the case where $\Delta$ is of order one.
We will find that the typical and maximum stopping distances decrease
for larger $\Delta$.

We will study the propagation of excitations of a massive
bulk field by studying the propagation of massive particles in
the bulk, similar to the massless case studied earlier in this
paper.  (An alternative discussion directly in terms of a wave
analysis is sketched in Appendix \ref{app:xmaxm}.)
We should emphasize that the term ``massive'' refers only to the
bulk fields and corresponding bulk ``particles'' in our discussion,
and so to the conformal dimension of the source operators in the
boundary theory.
We have not introduced any masses in the
4-dimensional strongly-coupled field theory: the
theory is still just ${\cal N}{=}4$ supersymmetric Yang Mills theory.


\subsection {Stopping distance of massive particles}
\label {sec:stopm}

For a particle of mass $m$ in 5 dimensions, the stopping distance integrals
(\ref{eq:null}) and (\ref{eq:stop1}) are modified to
\begin {equation}
   \xstop
   = \int \sqrt{g_{\five\five}} \, dx^\five \> 
      \frac{ g^{\mu\nu} q_\nu }
           { (-q_\alpha g^{\alpha\beta} q_\beta-m^2)^{1/2} }
   =
     \int
     \frac{du}{\sqrt{u(u^2-\frac{q^2}{|\q|^2})-\frac{(Rm)^2}{4|\q|^2} f }}
   \,.
\label {eq:stopm}
\end {equation}
As in the massless case, we shall see below that the stopping distance
will be dominated by $u \sim u_\star \ll 1$.  So we will be able
to approximate
$f\simeq 1$ above:
\begin {equation}
   \xstop
   \simeq
     \int
     \frac{du}{\sqrt{u(u^2-\frac{q^2}{|\q|^2})-\frac{(Rm)^2}{4|\q|^2}}}
   \,.
\label {eq:mstop}
\end {equation}

For $\Delta > d$ (in which case $m^2$ is positive), there is an
issue with the lower limit of integration in (\ref{eq:mstop}):
our classical
particle with 4-momentum $q_\mu$
cannot exist in the region where the square root
in (\ref{eq:mstop}) is imaginary.
For $u \ll u_*$ (and focusing on $q^2 < 0$), this condition
allows for a classical particle when
$u \ge u_{\rm min}$ with
\begin {equation}
   u_{\rm min} \simeq \frac{(Rm)^2}{-4q^2} \,.
\label {eq:umin}
\end {equation}
How to interpret this?  The wave equation is not well described
by geometric optics near the turning point $u_{\rm min}$.
However, as long as $u_{\rm min} \ll u_\star$, the calculation
of the stopping distance will be dominated by much larger $u$
(where $m$ is ignorable),
and so we may still use the particle picture to approximate
\begin {equation}
   \xstop
   \simeq
     \int_{\sim u_{\rm min}}^{1}
     \frac{du}{\sqrt{u(u^2-\frac{q^2}{|\q|^2})-\frac14 (Rm)^2}}
   \simeq
     \int_{0}^{\infty}
     \frac{du}{\sqrt{u(u^2-\frac{q^2}{|\q|^2})}}
   \,,
\end {equation}
which is the same as the massless particle result (\ref{eq:intdx3}).
In this respect, the mass can be ignored.

What happens at $u \sim u_{\rm min}$ can be made more
concrete by returning to the wave problem and
looking at the solution to the massive scalar wave
equation
\begin {equation}
   \frac{1}{\sqrt{-g}} \partial_\five
        (\sqrt{-g} g^{\five\five} \partial_\five \Phi)
   = (q_\mu g^{\mu\nu} q_\nu + m^2) \Phi
\label{eq:scalarEOM}
\end {equation}
in the limit $u \ll u_\star \ll 1$.  In this limit, one is close
enough to the boundary that AdS$_5$-Schwarzschild is approximately
just AdS$_5$, and the equation becomes the zero-temperature wave
equation of a massive scalar in AdS$_5$.  The retarded solution to this
equation is
\begin {equation}
  \Phi \simeq
   {\cal N}_q \, \frac{i\pi}{\Gamma(\nu)}
   \bigl(\tfrac12 \sqrt{-q^2} \bigr)^\nu
   (4 u)^{d/4} \,
   H_\nu^{(1)}(\sqrt{-4uq^2}) ,
\label {eq:Phim}
\end {equation}
where $H_\nu^{(1)}$ is the Hankel function, $d{=}4$ is the
space-time dimension of the boundary theory,
${\cal N}_q$ is an overall
normalization, and
\begin {equation}
   \nu = \Delta - \frac{d}{2} \,.
\label {eq:nu}
\end {equation}
The solution behaves like
\begin {equation}
   \Phi \simeq {\cal N}_q z^{d-\Delta}
\label {eq:PhiB}
\end {equation}
in the boundary limit $z {\to} 0$, where $z \equiv 2\sqrt{u}$.
The divergence of (\ref{eq:PhiB}) as $z \to 0$ for $\Delta > d$
(i.e.\ $m^2 > 0$) reflects the renormalization required of the
corresponding operators in the 4-dimensional gauge theory.
In our discussion, we will be able to ignore the details of
holographic renormalization prescriptions and simply summarize
that (\ref{eq:Phim}) should approach $z^{d-\Delta} \, \phi_{\rm b}(q)$
as $z$ approaches the (regulated) boundary,
where $\phi_{\rm b}$ is the (renormalized) boundary source.
Choosing
$\phi_{\rm b}(q) = 1$ defines the bulk-to-boundary propagator,
which corresponds to (\ref{eq:Phim}) with ${\cal N}_q \simeq 1$.%
\footnote{
   In more detail,
   follow Ref.\ \cite{FR} and
   normalize the bulk-to-boundary propagator
   to be $\zB^{d-\Delta}$ at $z{=}\zB$, where $\zB$ is
   arbitrarily small.  Then
   \[
     {\cal N}_q = 
     \zB^{d-\Delta}
     \left[
       \frac{i\pi}{\Gamma(\nu)}
       \bigl(\tfrac12 \sqrt{-q^2} \bigr)^\nu
       (4 \uB)^{d/4} \,
       H_\nu^{(1)}(\sqrt{-4\uB q^2})
     \right]^{-1}
     = 1 + O(\zB) ,
   \]
   and one takes $\zB \to 0$ at the very end of the calculation.
   In yet more detail, a systematic method for holographic
   renormalization is described in
   Refs.\ \cite{holoren1,holoren2,skenderis}.
}

The Hankel function goes through many oscillations, and
so is well approximated by the geometric optics limit, when its
argument is large compared to {\it both}\/ 1 and $\nu$.
In our case, this condition is parametrically equivalent to
\begin {equation}
   u \gg u_{\rm match} \equiv \frac{\max(1,(Rm)^2)}{-q^2} ,
\label {eq:umatch}
\end {equation}
which may be also be written as
\begin {equation}
   u \gg \max\Bigl( \frac{1}{-q^2} \, , \, u_{\rm min} \Bigr) .
\end {equation}
This generalizes the condition $u \gg 1/(-q^2)$ previously discussed
for the massless case.  If we convolve (\ref{eq:Phim}) with a high-energy
source (\ref{eq:source1}), we will not be able to use the particle
approximation to figure out the details of what is happening at
$u \sim u_{\rm min}$, but we will be able to use it when the
resulting wave packet propagates to $u \sim u_\star$ provided
$u_* \gg u_{\rm match}$, and so
we may then use the particle approximation to calculate the stopping
distance.

Because we get the same stopping distance as for the massless case,
we can take over (\ref{eq:stop}):
\begin {equation}
   \xstop
   \simeq \frac{c}{\sqrt2} \left( \frac{|\q|^2}{-q^2} \right)^{1/4}
   \simeq \frac{c}{2} \left( \frac{E}{\del} \right)^{1/4}
   .
\label {eq:stop2}
\end {equation}
The geometric optics approximation at $u\sim u_\star$
(and so this result for the stopping distance) will
fail unless $u_{\rm match} \ll u_\star$.
Using (\ref{eq:ustar}) and (\ref{eq:umatch}),
that condition requires
\begin {equation}
   \del \gg \left( \frac{E}{\max(1,(Rm)^4)} \right)^{-1/3} .
\label {eq:delmcondition}
\end {equation}
One might suspect that the particle approximation breaks down at
the maximum possible stopping distance, in which case
(\ref{eq:stop2}) then gives that maximum to be
\begin {equation}
   \xmax \sim \left( \frac{E}{\max(1,Rm)} \right)^{1/3}
   \sim \left( \frac{E}{\Delta} \right)^{1/3} .
\label {eq:xmaxm}
\end {equation}
We give a more detailed argument for this result
in Appendix \ref{app:xmaxm}.

Eq.\ (\ref{eq:xmaxm}) implies that the maximum stopping distance
decreases as the conformal dimension $\Delta$ of the BPS source
operator is increased.  This qualitative feature is not novel to
the strongly-coupled theory: it is true for the {\it weakly}-coupled
theory as well.
For the BPS operators,
large $\Delta$ corresponds to an operator with roughly $\Delta$ powers
of scalar fields, such as
$\tr(\phi^\Delta)$, where $\phi$ is one
of the three complex scalar fields in the theory.
In weak coupling, if we inject total energy $E$ with such an operator,
it will typically generate $\Delta$ particles that each have energy
of order $E/\Delta$.
In weak coupling, the stopping distance of a
particle with energy $E$ scales as $E^{1/2}$ (up to logarithms),
and so the stopping distance of the $\Delta$ particles each with
energy $E/\Delta$ will scale as $\xmax \sim (E/\Delta)^{1/2}$.


\subsection {Generalizing the power-law tail}
\label {sec:Pm}

In this section, we investigate how the $(\xz)^{-9}$ power-law
tail in Fig.\ \ref{fig:stop} generalizes to other choices of
source operator.  We will take the source term in the gauge-theory action
to be of the form
\begin {equation}
   \int_x e^{i\bar k\cdot x} \, \Lambda_L(x) \, {\cal O}(x) ,
\label {eq:sourcem}
\end {equation}
where ${\cal O}(x)$ is a scalar BPS operator with
dimension $\Delta$.  We will see, however, that our result
also applies to the case of ${\cal O}$ being a transverse-polarized
R current, which was the case discussed in section \ref{sec:reproduce}.

Since the dependence (\ref{eq:stop2}) of the stopping distance on
$q^2$ is the same as in the massless case, the only significant
qualitative difference in the distribution of stopping distances
will come from the distribution ${\cal P}(q^2)$ of $q^2$ created by
the source operator.  The shape of this distribution is determined by
the dimension $\Delta$ of the source operator, as we now describe.
As discussed in Refs.\ \cite{HIM,adsjet}, temperature does not
affect the initial creation of the jet, and so we can simplify
the analysis by evaluating ${\cal P}(q^2)$ at zero temperature.
Consider the probability density in $q$ associated with a source
operator ${\cal O}$ acting on the vacuum:
\begin {multline}
   \sum_{\rm any}
   \langle {\rm any} | [{\cal O}(q)]^\dagger | {\rm vac}\rangle^*
   \langle {\rm any} | [{\cal O}(q')]^\dagger | {\rm vac}\rangle
\\
   =
   \langle {\rm vac} | {\cal O}(q) \, [{\cal O}(q')]^\dagger
                     | {\rm vac} \rangle
   \equiv
   i G_>(q) \, (2\pi)^d \delta^{(d)}(q-q') ,
\label {eq:Gg}
\end {multline}
where $G_>$ is the Wightman correlator of ${\cal O}$.  At zero
temperature, it is related to the spectral density $\rho$ of the
operator by%
\footnote{
  Alternatively, we could use the finite-temperature relation
  $i G_>(q) = [1+n(q^0)] \, \rho(q)$, where
  $n(\omega) = (e^{\beta\omega}-1)^{-1}$ is the Bose distribution,
  and then use the fact that $q^0 \simeq E \gg T$ in our problem.
}
\begin {equation}
   i G_>(q) = \theta(q^0) \, \rho(q) .
\end {equation}
The distribution of jet 4-momenta is therefore given by the
spectral density $\rho(q)$.  At zero temperature, Lorentz
and scaling invariance
allow us to use simple dimensional analysis to know how
$\rho$ scales with $q$:%
\footnote{
   To get the dimension of $\rho(q)$,
   use (\ref{eq:Gg}) and note that ${\cal O}(x)$ having dimension
   $\Delta$ means that the Fourier transform ${\cal O}(q)$ has
   dimension $\Delta-d$.
}
\begin {equation}
   \rho(q) \propto \theta(-q^2) \, (-q^2)^\nu ,
\label {eq:rhom}
\end {equation}
where $\nu = \Delta-\frac12 d$ as in (\ref{eq:nu}).
The $\theta(-q^2)$ appears because only sources with
time-like $q^2$ produce persistent excitations at zero temperature.

If ${\cal O}$ were a vector operator $V^\mu$, like an R current, the
dimensional analysis would be complicated by the fact that one
could get factors of $q^\mu$ associated with the vector index
(rather than only factors of the virtuality $q^2$).
However the transverse spatial components
$\q^\perp$ of $q^\mu$ vanish, by definition.  So this complication
does not arise for the transverse R current operator $j^\perp$
that we discussed earlier, and (\ref{eq:rhom}) can also be used in
that case.

So far, we have only looked at the operator ${\cal O}$ and not the
other factors in the source term (\ref{eq:sourcem}).  We can
rewrite (\ref{eq:sourcem}) in $q$ space as
\begin {equation}
   \int_q \tilde\Lambda_L^*(q-\bar k) \, {\cal O}(q) .
\end {equation}
Correspondingly attaching a factor of $\Lambda_L^*(q-\bar k)$ to each
${\cal O}(q)$ in (\ref{eq:Gg}), we get a probability distribution
for $q$ proportional to
\begin {equation}
   \theta(q^0) \, \rho(q) \, |\tilde\Lambda_L(q-\bar k)|^2
   \propto
   \theta(-q^2) \, (-q^2)^\nu |\tilde\Lambda_L(q-\bar k)|^2 .
\end {equation}
For a transverse-translation invariant source,
the relative probability distribution for creating a jet with a
given $q_+$, where $q^2 \simeq 4 E q_+$, is then
\begin {equation}
   {\cal P}_+(\bar q_+)
   =
   \frac{
     \theta(-\bar q_+) \, |\bar q_+|^\nu
     \int d\Delta q_- \>
     \bigl|\tilde\Lambda_L^{(2)}(\bar q_+,\Delta q_-)\bigr|^2
   }{
     \int dq_+ \> d\Delta q_- \>
     \theta(-q_+) \, |q_+|^\nu
     \bigl|\tilde\Lambda_L^{(2)}(q_+,\Delta q_-)\bigr|^2
   } \,,
\label {eq:Pplus}
\end {equation}
which generalizes (\ref{eq:Pplus0}).
We will repackage this as
\begin {equation}
   {\cal P}_+(q_+) = 
   \frac{2^{(3+\nu)/2}L}{\Gamma(\frac{1+\nu}2)}
   \, \theta(-q_+) \, |q_+ L|^\nu \, \Psi_\nu(q_+ L),
\label {eq:PplusPsi}
\end {equation}
where
\begin {equation}
   \Psi_\nu(\bar q_+ L) \equiv
   \frac{
      \Gamma(\frac{1+\nu}2)
      \int dq_- \> |\tilde\Lambda_L^{(2)}(\bar q_+,q_-)|^2
   }{
       2^{(3+\nu)/2}L \int dq_+ \> dq_- \> \theta(-q_+) \,
       |q_+ L|^\nu  \bigl|\tilde\Lambda_L^{(2)}(q_+,q_-)\bigr|^2
   }
\end {equation}
has been normalized so that $\Psi_\nu(0)=1$ in the case of a Gaussian
source envelope (\ref{eq:Genvelope}).

Following (\ref{eq:Probx3}), the distribution of stopping distances is
then
\begin {align}
   \Prob(\xz) &\simeq
   \int d\del \, {\cal P}_+(-\del) \,
   \delta\bigl(\xz - \xstop(\del)\bigr) ,
\nonumber\\
   &\simeq
   \frac{2^{(3+\nu)/2}L}{\Gamma(\frac{1+\nu}2)}
   \int_0^\infty d\del \, (\del L)^\nu \,
   \Psi_\nu(-L\del) \,
   \delta\Bigl(\xz - \tfrac{c}{2} \left(\tfrac{E}{\del}\right)^{1/4}\Bigr)
\nonumber\\
   &=
   \frac{16 (\sqrt2\,c^4 EL)^{1+\nu}}
        {\Gamma(\frac{1+\nu}2) \, (2\xz)^{5+4\nu}} \,
   \Psi_\nu\Bigl(- \frac{c^4 EL}{(2\xz)^4}\Bigr) .
\label {eq:Probxm}
\end {align}
So the power-law tail in Fig.\ \ref{fig:stop} has
generalized to $(\xz)^{-(5+4\nu)} = (\xz)^{3-4\Delta}$, as shown
in Fig.\ \ref{fig:stopDelta}.
For the transverse R current operator, $\Delta=3$, which
recovers our previous result of $(\xz)^{-9}$ in that case.


\subsection {Gaussian Source Envelope}

Throughout this paper, we have discussed two different scales
$\xtypical \sim (EL)^{1/4}$ and $\xmax \sim E^{1/3}$
characterizing the stopping
distance.  In (\ref{eq:xmaxm}), we generalized the latter to
$\xmax \sim (E/\Delta)^{1/3}$ for the case of large $\Delta$.
Now we will discuss the similar generalization of $\xtypical$.
In general, we will still have
\begin {equation}
   \xtypical \sim \left( \frac{E}{(-q_+)_{\rm typical}} \right)^{1/4} ,
\end {equation}
but the relation between the typical $q_+$ of jets and the source
envelope size $L$ for large $\Delta$
will depend on details of how the source envelope
$\tilde\Delta_L(q)$ falls off for large $q_+$.
That's because the probability distribution
(\ref{eq:Pplus}) for the $q_+$ of the jet involves a
competition between the $|\bar q_+|^\nu$ factor
which favors large $|q_+|$ and the $|\Lambda_L(\bar q_+,\Delta q_-)|^2$
factor which suppresses $|q_+| \gg L^{-1}$.  The typical value of
$q_+$ represents a balance between the two and will scale with
$\nu$.  For the sake of a concrete
example, we consider here the case of a Gaussian source
envelope (\ref{eq:Genvelope}).  In this case, the
function $\Psi_\nu(q_+ L)$ in (\ref{eq:PplusPsi}) and
(\ref{eq:Probxm}) is simply
\begin {equation}
   \Psi_\nu(q_+ L) = e^{-2(q_+L)^2} .
\end {equation}
The typical values of $q_+$ from the probability distribution
(\ref{eq:PplusPsi}) then scale as
\begin {equation}
   (-q_+)_{\rm typical}
   \sim \frac{\nu^{1/2}}{L} \sim \frac{\Delta^{1/2}}{L},
\end {equation}
corresponding to%
\footnote{
Alternatively, one could compute the average value of $\xstop$
directly from (\ref{eq:Probxm}), giving
$
   \xz_{\rm avg} =
   \frac{c\,\Gamma(\frac38+\frac\nu2)}{2^{7/8}\,\Gamma(\frac12+\frac\nu2)}
   \, (EL)^{1/4}
$.
}
\begin {equation}
   \xtypical \sim \left( \frac{E L}{\sqrt{\Delta}} \right)^{1/4} .
\end {equation}

\section {Conclusion}
\label {sec:conclusion}


The 5-dimensional particle picture provides a relatively easy way
of understanding (from the gravity side of the calculation)
the appearance of the scale $(EL)^{1/4}$ in
jets created by finite-size sources in
strongly coupled ${\cal N}{=}4$ super Yang Mills plasmas.
By thinking about sources with different types of momentum distributions,
such as Figs.\ \ref{fig:source}a and b, we have learned that it
is the range of $q^2$ of the source which determines the range
of stopping distances.  Making $|q^2|$ larger causes the jet to
stop sooner.  For any finite size $L$ of source, the uncertainty
principle implies that there will
be a spread in the components of $q$ of at least order $1/L$ and so
a spread in $q^2$ of at least order $E/L$.  As a result,
almost all of the jets produced will travel
distances $\lesssim (EL)^{1/4}$
[of order $(EL)^{1/4}$ in the case of Fig.\ \ref{fig:source}a and
$\ll (EL)^{1/4}$ in the case of Fig.\ \ref{fig:source}b].
Events where a jet travels further (up to $E^{1/3}$) will always
be rare if the source size $L$ is small compared to the
maximum stopping distance scale $E^{1/3}$.

This interpretation, based on the 5-dimensional particle
picture, provides an important clarification to our original
derivation of the average distribution of charge deposition
shown in Fig.\ \ref{fig:stop}.  This average includes an
average over all events.  From the original result, it was
unclear whether on not Fig.\ \ref{fig:stop} qualitatively
tracks how the jet deposits its energy, momentum, and so forth on an
{\it event-by-event}\ basis.  It might have been that every single
jet produced deposits some of its energy at $\xz \sim E^{1/3}$
and most of its energy at $\xz \sim (EL)^{1/4}$.
The success of the 5-dimensional particle interpretation, and
in particular the success of (\ref{eq:Pstop}), indicates
that Fig.\ \ref{fig:stop} instead reflects a probability distribution
for how far the jet travels, and each individual jet dumps its
energy and charge in a very localized region of $\xz$.
It should be possible to independently verify this conclusion by
calculating correlations of the charge deposition at different
distances, which we will leave to future work.

Our analysis of massive 5-dimensional fields indicated that the
maximum stopping distance $E^{1/3}$ depends on the type of operator
used to create the jet and that the distance decreases as the
conformal dimensions of that operator increases.  That is, the
maximum stopping distance depends on the type of high-energy
excitation created.
This may shed some light on a discrepancy between (i) the stopping
distances found here and in Refs.\ \cite{adsjet,HIM}, which
find $\xmax \sim E^{1/3}$, and (ii) those
based on the evolution of classical strings in 5 dimensions
\cite{GubserGluon,CheslerQuark},
which find the parametrically smaller result
$\xmax \sim (E/\sqrt{\lambda})^{1/3}$,
where $\lambda \equiv N_{\rm c} g^2$ is the large 't Hooft coupling.
The gauge theory states corresponding to
classical strings in the gravity dual may simply
be states that are much more easily stopped by the strongly-coupled
quark-gluon plasma than are the states created by the source operators
considered in this paper.
One could then ponder which (if either) might be more instructive for
lessons about the theory of real QCD plasmas.
In order to further clarify the differences between the two
approaches, it would be interesting
to find a 4-dimensional gauge-theory
description of a source that could be precisely linked through duality
to the 5-dimensional initial classical string configurations that
have been used to study jet quenching.


\begin{acknowledgments}

We thank Djordje Minic for suggesting that we examine the dependence
of jet stopping distances on dimension $d$.
We also thank Andreas Karsch for useful conversations.
This work was supported, in part, by the U.S. Department
of Energy under Grant No.~DE-FG02-97ER41027 and by a
Jeffress research grant, GF12334.

\end{acknowledgments}

\appendix


\section {Stopping distances in different dimensions}
\label {app:dim}

The only relevant difference between AdS$_5$-Schwarzschild space and
AdS$_{d+1}$-Schwarzschild space is that the blackening function
$f=1-u^2$ in the metric (\ref{eq:metric}) is replaced by
$f=1-u^{d/2}$ \cite{AdSdim}.
The stopping distance (\ref{eq:stop1}) then becomes
\begin {align}
   \xstop
   &= \int_0^1 \frac{du}{\sqrt{u(u^{d/2}-\frac{q^2}{|\q|^2})}}
   \simeq \int_0^\infty \frac{du}{\sqrt{u(u^{d/2}-\frac{q^2}{|\q|^2})}}
\nonumber\\
   &= \frac{c_d}{\sqrt2} \left( \frac{|\q|^2}{q^2} \right)^{(d-2)/2d}
   \simeq \frac{c_d}{\sqrt2} \left( \frac{E}{4\del} \right)^{(d-2)/2d}
   ,
\end {align}
where
$c_d = \sqrt2 \, B(1{+}\frac1d, \frac12{-}\frac1d)$,
$B$ is the Beta function,
and we have assumed $d>2$.
Taking $\del \sim 1/L$, the
dominant stopping distance (\ref{eq:ELfourth}) for
our original source of Fig.\ \ref{fig:source}a then generalizes
from $(EL)^{1/4}$ to
\begin {equation}
   \xtypical \sim (EL)^{(d-2)/2d} .
\label {eq:ELd}
\end {equation}
The quickest way to estimate the maximum stopping distance,
generalizing $E^{1/3}$, is to estimate when the stopping distance
(\ref{eq:ELd})
becomes as large as the source itself, as we did for $d{=}4$
in section \ref{sec:intro}.
The result is
\begin {equation}
   \xmax \sim E^{(d-2)/(d+2)} .
\label {eq:xmaxd}
\end {equation}

The last result can also be obtained from a wave analysis by
analyzing the poles of the retarded bulk-to-boundary propagator,
just as was done for $d{=}4$ in Ref.\ \cite{adsjet}.
The scale of the exponential decay in (\ref{eq:final2}) was determined
by the imaginary part of the propagator pole closest to the real axis.
Here we simply follow section 4.6.2 of Ref.\ \cite{adsjet},
generalizing to arbitrary $d$.  The massless field equation for
$A_\perp$ is
\begin {equation}
   \left[ \partial_u^2 - \frac{4 E q_+ - u^{d/2} E^2}{u} \right] A_\perp = 0
   .
\end {equation}
Changing variables to $U \equiv e^{-i2\pi/(d+2)} E^{4/(d+2)} u$,
the field equation becomes
\begin {equation}
   \left[ -\partial_U^2 
          + \left( U^{(d-2)/2} - \frac{a}{U} \right)
   \right] A_\perp = 0
\label {eq:Schro}
\end {equation}
where
\begin {equation}
   a \equiv 4 E^{(d-2)/(d+2)} e^{-i\pi d/(d+2)} q_+ .
\label {eq:a}
\end {equation}
Poles of the bulk-to-boundary propagator occur when the
Schr\"odinger-like equation (\ref{eq:Schro}) has a
zero-energy bound state that vanishes at the origin.
The smallest value of $a$ for which this occurs is $O(1)$,
from which (\ref{eq:a}) gives that the pole closest to
the origin has
\begin {equation}
   \Im q_+^{\rm pole} \sim E^{-(d-2)/(d+2)} .
\end {equation}
So the response to the source falls exponentially as
\begin {equation}
   |e^{i q_+^{\rm pole} x^+}|^2 \sim e^{-\kappa_d E^{-(d-2)/(d+2)} x^+}
\end {equation}
for large $x^+$, for some constant $\kappa_d$.
This behavior is consistent with (\ref{eq:xmaxd}).


\section {Null geodesic in AdS$_5$-Schwarzschild}
\label {app:null}

For the sake of keeping this paper self-contained, we give here a brief
derivation of (\ref{eq:null}).
A null geodesic has
\begin {equation}
   0 = (ds)^2 = dx^\mu \, g_{\mu\nu} \, dx^\nu + dx^\five \, g_{\five\five} \, dx^\five ,
\end {equation}
and so
\begin {equation}
   \frac{dx^\five}{d\lambda} =
      \frac{1}{\sqrt{g_{\five\five}}}
      \left[ - g_{\mu\nu} \, \frac{dx^\mu}{d\lambda} \,
                           \frac{dx^\nu}{d\lambda} \right]^{1/2} ,
\label {eq:null1}
\end {equation}
where we will take $\lambda$ to be any affine parameter for the
trajectory.  Because of 4-dimensional translation invariance,
\begin {equation}
   g_{\mu\nu} \, \frac{dx^\nu}{d\lambda}
\end {equation}
is conserved and proportional to $q_\mu$, so that
\begin {equation}
   \frac{dx^\mu}{d\lambda} \propto g^{\mu\nu} q_\nu.
\end {equation}
Dividing this equation by (\ref{eq:null1}) gives
\begin {equation}
   \frac{dx^\mu}{dx^\five} =
      \sqrt{g_{\five\five}} \,
      \frac{ g^{\mu\nu} q_\nu }
           { (-q_\alpha g^{\alpha\beta} q_\beta)^{1/2} }
   \,,
\end {equation}
which in turn gives (\ref{eq:null}).


\section {More on the geometric optics approximation}
\label {app:GO}

Here, we will go into a little more detail about the conditions for the
geometric optics approximation, for space-time backgrounds with 4-dimensional
Poincar\'e invariance.  For the sake of concreteness, we will
consider the case of the source (\ref{eq:Lsource}) used in our earlier
work, where the source operator is a transverse-polarized R current and
so is dual to a 5-dimensional transverse vector field
$A_\perp = \pol_\mu A^\mu$ in
the gravity description.  The equation of motion for $A_\perp(q,x^\five)$,
where $q$ is the 4-momentum, has the form
\begin {equation}
   \frac{1}{\sqrt{-g}} \, 
   \partial_\five(\sqrt{-g} \,
        g^{\perp\perp} g^{\five\five} \partial_\five A_\perp)
   = g^{\perp\perp} q_\mu g^{\mu\nu} q_\nu A_\perp
\end {equation}
where
$g^{\perp\perp}$ is the component of the inverse metric in the direction
of the polarization, e.g.\ $g^{\perp\perp} = g^{11} = g^{22}$ for
$\q$ in the $\xz$ direction.
Now switch to coordinate
\begin {equation}
   \ell \equiv \int \sqrt{g_{\five\five}} \, dx^\five ,
\end {equation}
which parametrizes proper length in the direction of the fifth
dimension, and note for future reference that
\begin {equation}
   \partial_\ell = \frac{1}{\sqrt{g_{\five\five}}} \, \partial_\five
   .
\label {eq:Dell}
\end {equation}
The equation of motion is then
\begin {equation}
   w^{-1} \partial_\ell (w \partial_\ell A_\perp)
   = q_\mu g^{\mu\nu} q_\nu A_\perp .
\end {equation}
where
\begin {equation}
   w = \sqrt{-g_{(4)}} \, g^{\perp\perp}
\end {equation}
and $g_{(4)} = g^{\five\five} g$
is the determinant of the 4-dimensional part of the
5-dimensional metric.
(As another simple example, one could consider
the case of a source operator dual to a massless scalar field, which
would corresponds instead to taking $w = \sqrt{-g_{(4)}}$ in the equation
of motion.)  Now define
\begin {equation}
   a \equiv \sqrt{w} \, A_\perp
\end {equation}
to get
\begin {equation}
   \partial_\ell^2  a
   = \left[
       q_\mu g^{\mu\nu} q_\nu
       + \frac{1}{2\sqrt{w}} \, \partial_\ell
            \Bigl(\frac{\partial_\ell w}{\sqrt{w}}\Bigr)
   \right] a .
\end {equation}
This looks just like a one-dimensional quantum mechanics problem in
$\ell$ with wavenumber $k(\ell) \equiv \sqrt{2m(E-V(\ell))}$ replaced by
\begin {equation}
   k(\ell) =
   \sqrt{
       - q_\mu g^{\mu\nu} q_\nu
       - \frac{1}{2\sqrt{w}} \,
         \partial_\ell\Bigl(\frac{\partial_\ell w}{\sqrt{w}}\Bigr)
   } .
\label {eq:k}
\end {equation}
The WKB condition that the wavelength $\lambda(\ell) = 2\pi/k(\ell)$ in such
a quantum mechanics problem not change significantly over distances of one
wavelength is $\partial_\ell \lambda(\ell) \ll 1$.

For the metric (\ref{eq:metric}), $w = (R/2)^2 f^{1/2} u^{-1}$, and
(\ref{eq:k}) becomes
\begin {equation}
   k(u) =
   \frac{2}{R}
   \sqrt{
       \frac{u}{f}(u^2\q^2-q^2)
       +\frac{1-2u^2}{4f}
   }
\end {equation}
with the help of (\ref{eq:Dell}).
For the $u$'s of interest to our discussion of particles, which are
$u \sim u_\star \ll 1$, the $(1-2u^2)/4f$ term under the
square root is negligible, giving
\begin {equation}
   k(u) \simeq
   \sqrt{
       - q_\mu g^{\mu\nu} q_\nu
   }
   \simeq
   \frac{2}{R}
   \sqrt{
       u (u^2\q^2-q^2)
   } .
\end {equation}
Again using (\ref{eq:Dell}), the condition
$\partial_\ell \lambda(\ell) = \partial_\ell(2\pi/k) \ll 1$
then gives (\ref{eq:GOconditionu}) for $u \sim u_\star$.

\section {Maximum Stopping Distance for High-Dimension
              Source Operators}
\label {app:xmaxm}

In section \ref{sec:stopm}, we saw that the particle picture breaks down
when $\xz \gtrsim \xmax$ with
\begin {equation}
   \xmax \sim \left( \frac{E}{\max(1,Rm)} \right)^{1/3}
   \sim \left( \frac{E}{\Delta} \right)^{1/3} ,
\label {eq:xmaxm2}
\end {equation}
and we suggested that this $\xmax$ was the furthest that jets
would propagate---that is, that energy or charge deposition at
larger distances would be exponentially suppressed, similar to
our previous $\Delta{=}3$ result of (\ref{eq:final2}).
In this appendix, we will discuss how (\ref{eq:xmaxm2}) arises
in a wave analysis following the methods of Ref.\ \cite{adsjet}.
We will focus on the case of $\Delta \gg 1$.
As in the main text, we assume that the large $E$ limit is taken
{\it first}, and only then do we consider large $\Delta$.

There is a subtlety to the results we will find.
In section \ref{sec:xmaxpole},
we will analyze the
exponential fall-off of jet charge deposition at very large $x$
by finding the location in
the complex $q_+$ plane corresponding to the first quasi-normal
mode of the bulk field.  For large $\Delta$,
we will find an exponential fall-off
of the form
\begin {equation}
   \exp\left(- \frac{2c'_1 \xz}{E^{1/3}/\Delta^{4/3}}\right) ,
\label {eq:expm}
\end {equation}
where $c'_1$ is a constant.
This is the large-$\Delta$ version of the exponential in
(\ref{eq:final2}).
The $\xz$ scale that determines the rate of exponential fall-off
in (\ref{eq:expm}) is $E^{1/3}/\Delta^{4/3}$.
One might naturally guess that exponential suppression therefore
applies whenever $\xz \gg E^{1/3}/\Delta^{4/3}$ and so guess that
$\xmax \sim E^{1/3}/\Delta^{4/3}$ instead of (\ref{eq:xmaxm2}).
This guess fails, however, for reasons we shall now outline.


\subsection {Overview}
\label{sec:xmaxoverview}

\begin {figure}
\begin {center}
  \includegraphics[scale=0.4]{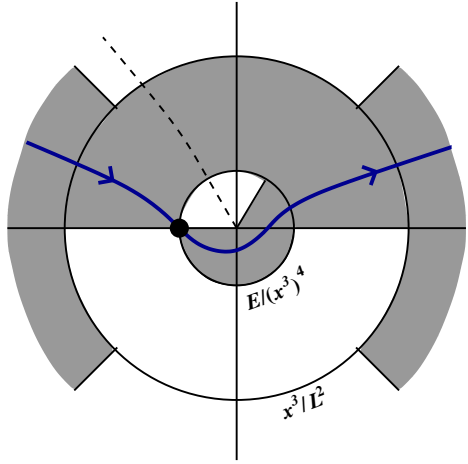}
  \caption{
     \label{fig:contour}
     Integration contour in the $q_+$ complex plane
     for saddle point approximations to the $q_+$ integral
     in (\ref{eq:calA}).  This is a slightly simplified version,
     appropriate for $u{\to} 1$,
     of Fig.\ 13b of Ref.\ \cite{adsjet}.
     The location of the saddle point $q_+^\star$ is
     marked by the large dot.  The circles indicate
     different parametric scales for $|q_+|$.
  }
\end {center}
\end {figure}

To understand the issues involved, we briefly highlight
some relevant aspects of the
$m{=}0$ calculation from Ref.\ \cite{adsjet}.
The main part of the calculation there involved computing the bulk
response ${\cal A}$ to the high-energy source on the boundary, given
by
\begin {equation}
  \Field(x,u)
  \equiv
  \int_q 
  \calGRup(q,u) \,
  \tilde\Lambda_L(q-\kbig) \,
  e^{iq\cdot x} ,
\label {eq:calA}
\end {equation}
where ${\cal G}$ is the bulk-to-boundary propagator.
Our result for the stopping distance came from
extracting the behavior of $\Field$ near the horizon,
$u\to1$.%
\footnote{
   Readers may wonder at the connection between (i) studying
   $u\to1$ and (ii) studying $u\sim u_\star \ll 1$ as in
   the particle arguments in the main text of this paper.
   The point is that how far the particle travels is determined
   by where it is when $u\to1$, but the integral which gives
   that distance is dominated by $u\sim u_\star \ll 1$.
   In the wave analysis of Ref.\ \cite{adsjet}, we studied
   the response at late times,
   corresponding to $u{\to}1$ for the bulk excitation.
   Our results for the near-horizon bulk response were determined
   by the $E^{1/4} (-q_+)^{3/4}$
   term in the WKB exponent $S$ for ${\cal G}$ (see eq.\ (4.51)
   of Ref.\ \cite{adsjet}).  But this term was generated by
   the $u {\sim} u_\star$ region of the integral that gave $S$.
   See, for example, eq.\ (D13) of Ref.\ \cite{adsjet}, which
   is proportional to the current paper's particle stopping distance
   integral (\ref{eq:stop1}).
}
The critical part of the $q$ integration was the integral
over $q_+$.  For $\xz \ll E^{1/3}$ (in the $m{=}0$ case),
we found that we could
deform the $q_+$ integration
contour in the complex plane so that the integral was everywhere
exponentially suppressed except at a saddle point%
\footnote{
   In Ref.\ \cite{adsjet}, we expressed formulas in terms
   of $X^+ \equiv x^+ - \tau(u)$ instead of $\xz$.  As
   discussed in that paper, the late-time response is localized
   to $x^- \simeq -\tau(u)$ (see eq.\ (4.49) of Ref.\ \cite{adsjet}),
   and so $X^+ \simeq 2 \xz$.
}
\begin {equation}
   q_+^\star \simeq - \frac{c^4 E}{(2\xz)^4}
\label {eq:qstar}
\end {equation}
of (\ref{eq:wave}).
This contour is depicted in Fig.\ \ref{fig:contour}.
Parametrically far into the interior of the shaded region indicates places
where the magnitude of the integrand is exponentially suppressed.
Parametrically far into the unshaded regions indicates places where
it is exponentially large.  The dashed line depicts
a line of poles of the bulk-to-boundary propagator $\calGRup$,
corresponding to quasi-normal modes.
In the WKB approximation to that propagator, this line of poles became
a cut.

\begin {figure}
\begin {center}
  \includegraphics[scale=0.4]{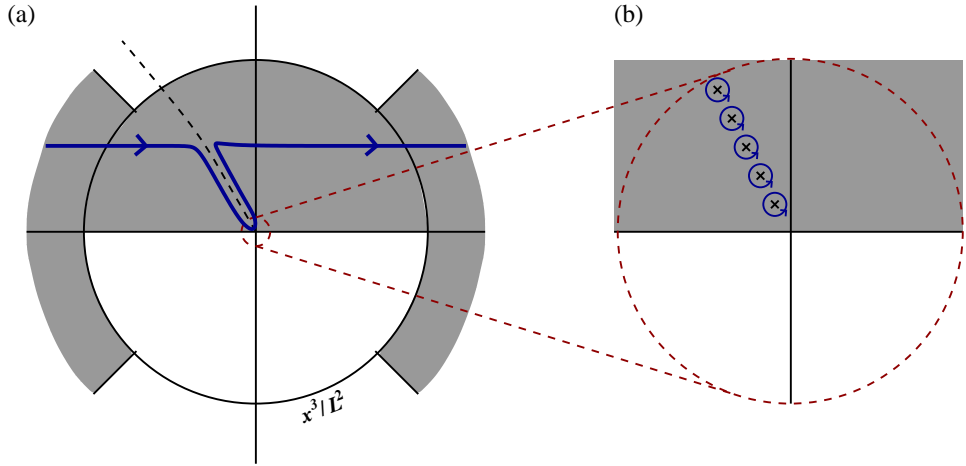}
  \caption{
     \label{fig:contour2}
     (a) Similar to Fig.\ \ref{fig:contour} but for the case
     $\xz \gg \xmax$.
     (b) A magnification of the region near the origin.
  }
\end {center}
\end {figure}

For $\xz \gg E^{1/3}$, the regions of exponential suppression
for the integrand are shown in 
Fig.\ \ref{fig:contour2}a.  Saddle point methods are unreliable.
Instead, choose the integration contour
shown there.  The piece that goes around the line of poles can
be re-expressed as a sum of contributions from each pole, as shown
in Fig.\ \ref{fig:contour2}b, which depicts a magnification of
the neighborhood of the origin of Fig.\ \ref{fig:contour2}a.
The $e^{i q_+\cdot x^+}$ piece of the $e^{iq\cdot x}$ in
(\ref{eq:calA}) causes the contributions from the poles to
be exponentially suppressed according to their distance
$\Im q_+$ from the real axis.  For $\xz \gg E^{1/3}$, the
nearest pole dominates and produces the exponential fall-off
(\ref{eq:final2}) of the jet's charge deposition.

\begin {figure}
\begin {center}
  \includegraphics[scale=0.4]{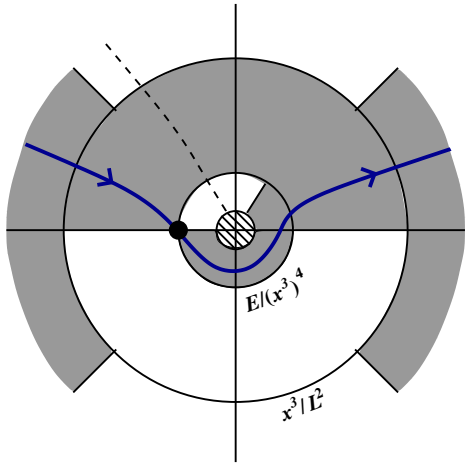}
  \caption{
     \label{fig:contour3}
     Like Fig.\ \ref{fig:contour} but for the case of large
     $\Delta$ and $\xz \ll E^{1/3}/\Delta^{1/3}$.
     The cross-hatched region represents
     $|q_+| \lesssim \Delta^{4/3}/E^{1/3}$.
  }
\end {center}
\end {figure}

Now we return to the massive case, with $\Delta \gg 1$, and ask
what happens for
\begin {equation}
   \frac{E^{1/3}}{\Delta^{4/3}} \ll \xz \ll \frac{E^{1/3}}{\Delta^{1/3}}
   .
\label {eq:xzrange}
\end {equation}
This is the interesting case, where
(i) the geometric optics and particle arguments of
section \ref{sec:massive} indicate that the charge deposition is not
exponentially suppressed
but (ii) the guess we might make based on
(\ref{eq:expm}) suggested that it is suppressed.
As we shall discuss in section \ref{sec:xmaxsaddle},
the mass does not significantly affect the massless picture
of Fig.\ \ref{fig:contour} provided
\begin {equation}
   |q_+| \gg \frac{\Delta^{4/3}}{E^{1/3}} ,
\label {eq:qcondition}
\end {equation}
which is the condition (\ref{eq:delmcondition}) discussed in
the main text.  For $\xz \ll E^{1/3}/\Delta^{1/3}$ as
in (\ref{eq:xzrange}), the condition (\ref{eq:qcondition})
is satisfied at the
saddle point (\ref{eq:qstar}) and for larger $|q_+|$.  So, as long
as we are careful to route the contour as in Fig.\ \ref{fig:contour3},
we can take over the methods of the $m{=}0$ calculation, make
a saddle point approximation to the $q_+$ integral
(which corresponds to making the geometric optics approximation),
and so find a result that is not exponentially suppressed.

\begin {figure}
\begin {center}
  \includegraphics[scale=0.4]{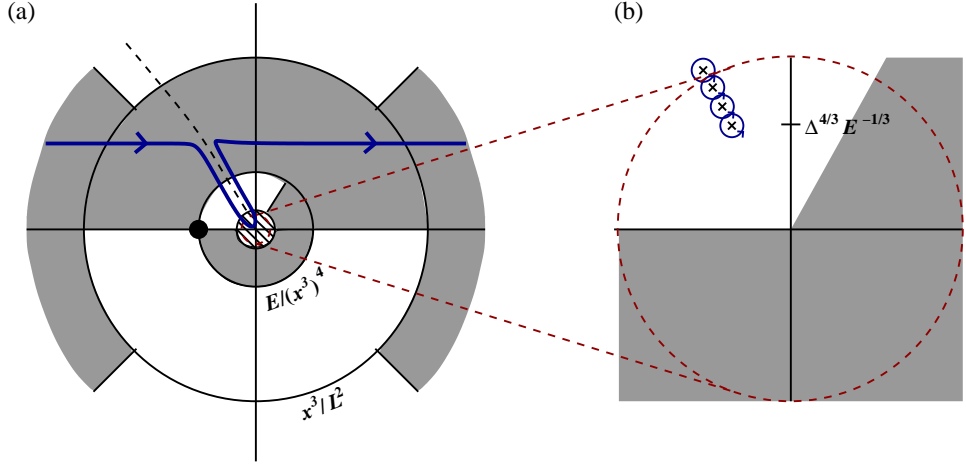}
  \caption{
     \label{fig:contour4}
     Like Fig.\ \ref{fig:contour3} but using a contour
     that picks up the poles instead of passing through
     the saddle point.
  }
\end {center}
\end {figure}

But now consider a large $\xz$ calculation, along the lines of
Fig.\ \ref{fig:contour2}.  We shall see in section
\ref{sec:xmaxpole} that the closest pole to the real axis has
\begin {equation}
   \Im q_+^{\rm pole} \sim \frac{\Delta^{4/3}}{E^{1/3}} .
\end {equation}
That is, the line of poles in Fig.\ \ref{fig:contour3} begins at
the edge of the hatched circle.  For $\xz$ in the range of
(\ref{eq:xzrange}), picking up the poles is not so useful.
Fig.\ \ref{fig:contour4} shows the large-$\Delta$ analog
of Fig.\ \ref{fig:contour2} for this $\xz$ range.  As one
looks at poles progressively further from the origin, the
$\exp(iq\cdot x)$ factors leads to suppression, as before,
but the $\calGRup$ factor {\it grows}\/ exponentially (as
well as oscillates), and this exponential growth dominates
the integrand in
the unshaded region of Fig.\ \ref{fig:contour4}.  So one cannot
approximate the integral by the contribution from the nearest pole,
and asymptotic formulas like (\ref{eq:expm}) do not apply in this case.

Since the saddle point approximation is controlled and gives a
result that is not exponentially suppressed in the range
(\ref{eq:xzrange}), whereas the approximations that lead to
(\ref{eq:expm}) are not valid there, we conclude that the maximum
stopping distance is of order $E^{1/3}/\Delta^{1/3}$ and not
$E^{1/3}/\Delta^{4/3}$.

\begin {figure}
\begin {center}
  \includegraphics[scale=0.4]{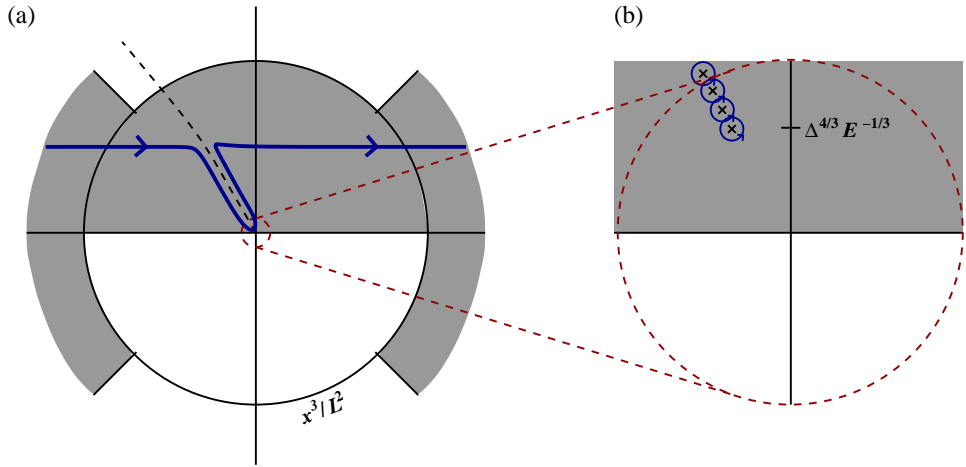}
  \caption{
     \label{fig:contour5}
     Like Fig.\ \ref{fig:contour4} but for the case
     $\xz \gg E^{1/3}/\Delta^{1/3}$.
  }
\end {center}
\end {figure}

What happens as one continues to increase $\xz$?  For
$\xz \gg E^{1/3}/\Delta^{1/3}$, the saddle point approximation
breaks down and Fig.\ \ref{fig:contour4} becomes
Fig.\ \ref{fig:contour5}.  In this case, the calculation is
dominated by the first pole and (\ref{eq:expm}) applies.
We will not attempt here to calculate the details of the transitional
behavior at $x \sim \xmax$.%
\footnote{
   However, for the sake of showing that
   (\ref{eq:expm}) is mathematically consistent with
   $\xmax \sim E^{1/3}/\Delta^{1/3}$, we point out that
   $\{ \exp[ \frac{2c'_1 (x-\xmax)}{E^{1/3} \Delta^{-4/3}} ] + 1 \}^{-1}$
   is an example of a function that is unsuppressed for
   $x \ll \xmax$ but decays like (\ref{eq:expm}) for
   $x \gg \xmax$.
}

\subsection{Quasi-normal modes}
\label {sec:xmaxpole}

To find the poles of $\calGRup$ in the complex $q_+$ plane for
large $E$, we follow the method of Ref.\ \cite{adsjet}.  For the
sake of simplicity, we will focus on the case of a massive bulk
scalar field, whose equation of motion is (\ref{eq:scalarEOM}).
As discussed in Ref.\ \cite{adsjet}, the pole positions at high
energy are determined (up to small corrections)
by the nature of the equation of motion for $u \ll 1$.
It will be more convenient to work with the variable
$z = 2\sqrt{u}$ instead of $u$.
Writing
\begin {equation}
   \phi = z^{(d-1)/2} \psi ,
\end {equation}
the equation of motion for $u\ll 1$ becomes the
Schr\"odinger-like equation
\begin {equation}
   -\tfrac12 \partial_z^2\psi 
   + V(z) \psi = -\tfrac12 q^2 \psi
\end {equation}
with potential
\begin {equation}
   V(z)
   = \frac12\left[
       -\left( \frac{z}{2} \right)^{d} \q^2
       + \frac{(Rm)^2+\frac{d^2-1}{4}}{z^2}
     \right] .
\end {equation}
Taking the high energy limit and
and setting $d{=}4$, this is
\begin {equation}
   -\tfrac12 \partial_z^2\psi 
   + V(z) \psi \simeq -2 E q_+ \psi
\end {equation}
with
\begin {equation}
   V(z)
   \simeq
   \frac12 \left[
       -\left( \frac{z}{2} \right)^{4} E^2
       + \frac{{\cal M}^2}{z^2}
   \right]
\end {equation}
and
\begin {equation}
   {\cal M}^2 \equiv (Rm)^2 + \tfrac{15}{4} \,.
\end {equation}
Following Ref.\ \cite{adsjet}, make the change of variables
from $u$ to
\begin {equation}
   U \equiv e^{-i\pi/3} E^{2/3} u ,
\end {equation}
which turns the retarded boundary condition at large $u$ into
the requirement that $\psi$ be real and exponentially falling.
In terms of $z$, this redefinition is
\begin {equation}
   Z \equiv e^{-i\pi/6} E^{1/3} z .
\end {equation}
The resulting equation is%
\footnote{
  For comparison with Ref.\ \cite{adsjet}, one may write a similar
  equation in terms of $U$ by defining $\phi = u^{(d-2)/4} \bar\phi$,
  giving
  $[ - \partial_U^2 + U + \frac{(Rm)^2+3}{4U^2} - \frac{a}{U} ]
   \bar\phi = 0$
  for $d=4$.
  This reduces to (4.66) of Ref.\ \cite{adsjet}
  for $A_\perp$ in the case $(Rm)^2=-3$, corresponding to $\Delta=3$.
  The formulation in the current paper in terms of $Z$ is more
  convenient because the pole locations $q_+$ can be identified as proportional
  to the bound state energies of a Schr\"odinger potential ${\cal V}(Z)$.
  Note also that ${\cal M}^2$ plays a roll analogous to angular momentum
  squared in the Schr\"odinger problem (\ref{eq:Schro2}), with
  ${\cal M}^2/Z^2$ like a centrifugal potential and the large-${\cal M}$
  limit analogous to a large angular momentum limit.
}
\begin {equation}
   -\tfrac12 \partial_Z^2\psi 
   + {\cal V}(Z) \psi = \tfrac12 a \psi
\label {eq:Schro2}
\end {equation}
with
\begin {equation}
   {\cal V}(Z)
   = \frac12 \left[
       \left( \frac{Z}{2} \right)^{4}
       + \frac{{\cal M}^2}{Z^2}
     \right]
\end {equation}
and $a$ defined in terms of $q_+$ as in Ref.\ \cite{adsjet}:
\begin {equation}
   q_+ = \tfrac14 \, e^{i 2\pi/3} E^{-1/3} a .
\label {eq:adef}
\end {equation}
Solving (\ref{eq:Schro2}) with the desired boundary conditions is
equivalent to setting $a$ to be twice the bound-state energies
associated with the potential ${\cal V}(z)$.
For ${\cal M}\gg 1$, these can be well approximated by treating
${\cal V}(z)$ in harmonic-oscillator approximation around its
minimum.  The result is
\begin {equation}
   \tfrac12 \, a_n
   = \tfrac38 {\cal M}^{4/3}
     + (n+\tfrac12) \sqrt{\tfrac32} \, {\cal M}^{1/3}
     + O({\cal M}^{-2/3})
\end {equation}
for $n=0,1,2,\cdots$.%
\footnote{
   Here we label the first pole $a_0$.
   In Ref.\ \cite{adsjet} we instead called it $a_1$.
}
Using (\ref{eq:adef}), we find that the first pole in $q_+$
is a distance of order $\Delta^{4/3}/E^{1/3}$ from the real axis,
but the spacing between successive poles in Fig.\ \ref{fig:contour5}b
is only of order $\Delta^{1/3}/E^{1/3}$.  The specific result
for $a_0$ determines
\begin {equation}
   c'_1 = \frac{\sqrt3 \, a_0}{4 \Delta^{4/3}} \simeq \frac{3\sqrt3}{16}
\end {equation}
for the exponential fall-off (\ref{eq:expm}) in the case $\Delta \gg 1$.


\subsection{Saddle point analysis}
\label {sec:xmaxsaddle}

In section \ref{sec:xmaxoverview}, we claimed that a large mass $m$ for the
bulk field does not qualitatively change the massless saddle-point
picture of Fig.\ \ref{fig:contour} except inside the (avoidable)
hatched region of Fig.\ \ref{fig:contour3}.  Here, we will briefly
sketch why.
For $Rm \gg 1$,
the condition for the validity of the WKB approximation%
\footnote{
   Essentially: that the derivative of
   the WKB exponent does not change significantly over one
   e-folding or oscillation.
}
is satisfied in the
small $u\to 0$ regime $u \ll u_{\rm min}$ as well as in the oscillatory
regime $u \gg u_{\rm min}$.
(The turning point $u \sim u_{\rm min}$ can be avoided simply by
analytically continuing around it, as in the textbook discussion of
WKB in Ref.\ \cite{LLQM}).  So we may use WKB all the way to the
boundary $u = \uB$:
\begin {equation}
   \calGRup \propto
   e^{iS} \equiv
   \exp\left[ i \int_{\uB}^u  du' \> q_\five(u') \right]
\label {eq:calS}
\end {equation}
where, for the sake of simplicity of presentation, we
will suppress showing the WKB prefactor.  For the massive
case,
\begin {equation}
   q_\five(u) =
   \frac{1}{f}
   \sqrt{ \frac{{u}^2 |\q|^2-q^2}{u} - \frac{(Rm)^2 f}{4{u}^2} } .
\label {eq:q5m}
\end {equation}
The integrand in (\ref{eq:calA}) then has exponential dependence
\begin {equation}
   e^{iq\cdot x} \calGRup \propto e^{i{\cal S}}
   \equiv e^{i(q\cdot x + S)} ,
\end {equation}
and the saddle point of its integral is determined by
\begin {equation}
  0 = \frac{\partial{\cal S}}{\partial q_\mu}
  = \frac{\partial}{\partial q_\mu}
  \left[ q \cdot x + \int dx^\five\> q_\five(x^\five) \right] ,
\end {equation}
which gives
\begin {equation}
  x^\mu = - \int dx^\five \> \frac{\partial q_\five}{\partial q_\mu} .
\end {equation}
Together with (\ref{eq:q5m}), this simply reproduces the particle-based
formula (\ref{eq:stopm}) for the stopping distance.  Approximating
$q_- \simeq E$ and solving for $q_+$ in terms of $\xz$ will then
give the saddle point $q_+^\star$ for the $q_+$ integration in
(\ref{eq:calA}).
We've already discussed the effect of the mass on the particle
stopping formula (\ref{eq:stopm}) back in section \ref{sec:stopm}.
Tracing the discussion of section \ref{sec:stopm} backward,
$\xz \ll (E/\Delta)^{1/3}$ corresponds to $-q_+$ given by
$\epsilon \gg (E/\Delta^4)^{-1/3}$, which corresponds in turn
to $u_{\rm min} \ll u_\star$.  That's precisely the case where
the mass had a negligible effect on the relationship between
$\xz$ and $q_+$.  In consequence, the mass $m$ will not have a
significant effect on the determination of the saddle point
$q_+^*$ for $\xz \ll (E/\Delta)^{1/3}$.

What about the behavior of the integrand elsewhere along the
contour in Fig.\ \ref{fig:contour}?  A discussion of the WKB
exponent $S$ of (\ref{eq:calS}) is complicated by the
divergence (\ref{eq:PhiB}) of the bulk-to-boundary propagator
on the boundary, which shows up as a logarithmic divergence
($\propto \ln \uB$) of the integral in (\ref{eq:calS}).  We will briefly
indicate in section \ref{sec:xmaxdiv} how one can do a WKB
analysis that avoids this divergence, but such details lose the forest
for the trees.  More simply,
the $\ln\uB$ divergence of $\int du\> q_\five$
is independent of $q_+$ and so does not affect the $q_+$
dependence of the integrand in (\ref{eq:calA}), and so it
will only affect the result by overall factors.
To focus on the question of whether the mass makes a significant
effect on the $q_+$ dependence, look at the effect of the mass on
$\partial{\cal S}/\partial q_+$ instead of on ${\cal S}$.
So look at
\begin {equation}
   \frac{\partial{\cal S}}{\partial q_\mu}
   = x^\mu + \int dx^\five \> \frac{\partial q_\five}{\partial q_\mu} \,.
\end {equation}
The first term is mass independent, and the second term is just
once again our integral  for the particle stopping distance as
a function of $q_+$, given by the right-hand side of (\ref{eq:stopm}),
though with an imaginary part even for real
negative $q_+$ due to integrating over $u < u_{\rm min}$.
However, if $|q_+| \gg \Delta^{4/3}/E^{1/3}$, then the effect of the
mass on this integral will have negligible relative magnitude, just
as in the previous discussion concerning the location of the
saddle point.


\subsection{Avoiding WKB exponent divergences}
\label {sec:xmaxdiv}

Finally, we sketch how one could set up a finite WKB
integral if one wanted to carry through the analysis of this
appendix in more detail than we have given.
To get the normalization (\ref{eq:PhiB}) appropriate for the
bulk-to-boundary propagator ${\cal G}$, we want $\Phi$ to give
$\uB^{(d-\Delta)/2}$ at $u{=}\uB$.  So
\begin {equation}
   \calGRup \approx
   \uB^{(d-\Delta)/2} \exp\left[ i \int_{\uB}^u  du' q_\five(u') \right]
   ,
\end {equation}
where we have again suppressed showing the WKB prefactor, other
than the overall power of $\uB$.
We can trade the divergent $\uB^{(d-\Delta)/2}$ normalization factor for
a finite $u^{(d-\Delta)/2}$ by introducing a compensating change in the
exponent:
\begin {equation}
   \calGRup
   \approx
   u^{(d-\Delta)/2}
   \exp\left[ i \int_{\uB}^u  du'
      \Bigl( q_\five(u') - i \frac{(\Delta-d)}{2u'} \Bigr)
   \right] .
\end {equation}
In the large $\Delta$ limit (required for our WKB analysis in
the region $u \ll u_{\rm min}$),
$\Delta-d \simeq Rm$, and so
we will replace the last equation by%
\footnote{
  We are sweeping something under the rug here.  One makes a small
  relative error in replacing $\Delta-d$ by $Rm$ when $\Delta$ is large,
  but in exponents
  one should really focus on absolute rather than relative
  errors.  We leave further
  refinement to the interested reader.
}
\begin {equation}
   \calGRup
   \approx
   u^{(d-\Delta)/2}
   \exp\left[ i \int_{\uB}^u  du' \left\{
        \frac{1}{f(u')}
        \sqrt{ \frac{{u'}^2 |\q|^2-q^2}{u'} - \frac{(Rm)^2 f(u')}{4{u'}^2} }
        - i \frac{Rm}{2u'}
     \right\} \right] .
\end {equation}
Now the integral in the exponent is finite if we take the limit
$\uB \to 0$, and so the appropriate WKB expression (still suppressing
showing the original WKB prefactor) can be approximated as
\begin {equation}
   \calGRup
   \approx
   u^{(d-\Delta)/2}
   \exp\left[ i \int_{0}^u  du' \left\{
        \frac{1}{f(u')}
        \sqrt{ \frac{{u'}^2 |\q|^2-q^2}{u'} - \frac{(Rm)^2 f(u')}{4{u'}^2} }
        - i \frac{Rm}{2u'}
     \right\} \right] .
\end {equation}
One may then use this WKB formula to pursue a more detailed analysis.
The $-i Rm/2u'$ subtraction in the integral cancels the original
integrand for $u \ll u_{\rm min}$ and so keeps the integral finite.
Its contribution for $u \gg u_{\rm min}$ will introduce a
additive piece of approximately
$\frac{i}2 Rm \ln u_{\rm min} \simeq \frac{i}2 \Delta\ln u_{\rm min}$
in the WKB exponent $S$.  That corresponds to
a multiplicative factor of $u_{\rm min}^{-\Delta/2}$ in the result for
$e^{i{\cal S}}$ and so $u_{\rm min}^{-\Delta}$ in $|{\cal A}|^2$.
Using (\ref{eq:umin}) for $u_{\rm min}$
and then the saddle-point value
(\ref{eq:qstar}) for $q_+$, this factor is
\begin {equation}
   u_{\rm min}^{-\Delta} \propto (q^2)^\Delta \propto 
   (q_+^\star)^\Delta \propto (\xz)^{-4\Delta} ,
\end {equation}
which is just the dependence of the power-law
tail on $\nu \simeq \Delta$ that we previously found in (\ref{eq:Probxm}).




\end {document}